\newif\ifLIPICS
\LIPICStrue

\documentclass[thm-restate, a4paper]{lipics-v2021}
\usepackage{amssymb,amsmath,bbold,amsthm,appendix}
\usepackage{qtree} %
\usepackage{hyperref,placeins}
\hypersetup{
    colorlinks = true,
    linkcolor = blue,
    urlcolor = blue,
    citecolor = blue,
}
\usepackage{multirow}
\usepackage[table]{xcolor}
\usepackage{graphicx,xspace,graphbox,tikz}
\usepackage{algorithm}
\usepackage[noend]{algpseudocode}
\usepackage{algorithmicx}
\usepackage[at]{easylist}
\usepackage{fancybox}
\usepackage{tikz}
\usepackage{caption}

\newcommand{\continueComment}[1]{
    \algrenewcommand{\algorithmiccomment}[1]{\hfill #1}
    \Comment{#1}
    \algrenewcommand{\algorithmiccomment}[1]{\hfill \(\triangleright\) #1}}

\usepackage{bm}

\newcommand{\algnamestyle}[1]{{\textsf{#1}}\xspace}
\newcommand{\TopDown}{\algnamestyle{TopDown}}
\newcommand{\Toy}{\algnamestyle{ToyDown}}

\newcommand{\Greedy}{\algnamestyle{Greedy}}
\newcommand{\Square}{\algnamestyle{Square}}
\newcommand{\ReCom}{\algnamestyle{ReCom}}
\newcommand{\Disconn}{\algnamestyle{Disconn}}

\newcommand{\frag}{\mathsf{Frag}}

\newcommand{\sdot}{\! \cdot \!}

\DeclareMathOperator{\E}{\mathbb{E}}
\renewcommand{\Pr}{\mathbb{P}}
\DeclareMathOperator{\Var}{Var}
\newcommand{\eps}{\varepsilon}
\newcommand{\ahat}{\widehat{a}}
\newcommand{\Lap}{\mathrm{Lap}}

\newcommand{\bbN}{\mathbb{N}}

\newcommand{\R}{\mathbb{R}}

\newcommand{\fragde}[1]{}
\newcommand{\floor}[1]{\lfloor #1 \rfloor}

\DeclareMathOperator*{\argmin}{arg\,min}

\title{Census TopDown: The Impacts of Differential Privacy on Redistricting}

\funding{This project was supported on NSF OIA-1937095 (Convergence Accelerator) and by a grant from the Alfred P. Sloan Foundation.}
\author{Aloni Cohen}{Hariri Institute for Computing and School of Law, Boston University, USA}{aloni.cohen@gmail.com}{}{NSF CNS-1414119 and CNS-1915763; DARPA HR00112020021}
\author{Moon Duchin}{Department of Mathematics, Tufts University, USA}{mduchin@mggg.org}{}{NSF DMS-2005512}
\author{JN Matthews}{Tisch College of Civic Life, Tufts University, USA}{jnmatthews@mggg.org}{}{}
\author{Bhushan Suwal}{Tisch College of Civic Life, Tufts University, USA}{bhushan@mggg.org}{}{}

\date{February 15, 2021}

\titlerunning{Census TopDown: The Impacts of Differential Privacy on Redistricting} %
\authorrunning{A.\ Cohen, M.\ Duchin, JN Matthews, and B.\ Suwal}
\Copyright{Aloni Cohen, Moon Duchin, JN Matthews, and Bhushan Suwal}

\ccsdesc[500]{Security and privacy}
\ccsdesc[500]{Applied computing~Law}
\ccsdesc[500]{Applied computing~Voting / election technologies} %

\keywords{Census, \TopDown, differential privacy, redistricting, Voting Rights Act}
\supplement{\url{https://mggg.org/DP}}

\acknowledgements{Authors are listed alphabetically. We thank Denis Kazakov, Mark Hansen, and Peter Wayner.  Kazakov developed the reconstruction algorithm as a member of Hansen's research group. Wayner guided our deployment of \TopDown in AWS and was an invaluable team member for the technical report.
Any opinions, findings, and conclusions or recommendations expressed in
this material are those of the authors and do not necessarily reflect the views of our funders.}

\EventEditors{Katrina Ligett and Swati Gupta}
\EventNoEds{2}
\EventLongTitle{2nd Symposium on Foundations of Responsible Computing (FORC 2021)}
\EventShortTitle{FORC 2021}
\EventAcronym{FORC}
\EventYear{2021}
\EventDate{June 9--11, 2021}
\EventLocation{Virtual Conference}
\EventLogo{}
\SeriesVolume{192}
\ArticleNo{5}

\begin{document}

\maketitle

\begin{abstract}
The 2020 Decennial Census will be released with a new disclosure avoidance system in place, putting {\em differential privacy} in the spotlight for a wide range of data users.  We consider several key applications of Census data in redistricting, developing tools and demonstrations for practitioners who are concerned about the impacts of this new noising algorithm called \TopDown.  Based on a close look at reconstructed Texas data, we find reassuring evidence that \TopDown will not threaten the ability to produce districts with tolerable population balance or to detect signals of racial polarization for Voting Rights Act enforcement.
\end{abstract}

\section{Introduction}

A new disclosure avoidance system is coming to the Census: the 2020 Decennial Census releases will use an algorithm called \TopDown  to protect the data from increasingly feasible {\em reconstruction attacks} \cite{TopDown}.
Census data is structured in a nesting sequence  of geographic units covering the whole country, from nation at the top to small {\em census blocks} at the bottom.
\TopDown starts by setting a {\em privacy budget} $\eps>0$ which is allocated to the levels of a designated hierarchy, then adding noise at each level in a {\em differentially private} way \cite{DMNS}.
When $\eps\to\infty$, the data alterations vanish, while $\eps\to 0$ yields pure noise with no fidelity to the input data.  The algorithm continues with a post-processing step that leaves an output dataset that is designed to be suitable for public use.

{\em Redistricting} is the process of dividing a polity into territorially delimited pieces in which elections will be conducted.  The Census has a special release---named the PL 94-171 after the law that requires it---that reports the number of residents in every geographic unit in the country by race, ethnicity, and the number of voting-age residents \cite{PL94}.
The 2020 release is slated to occur by September 2021, after which many thousands of district lines will be redrawn: not only U.S.\ Congressional districts, but those for state legislatures, county commissions, city councils, and many more.

Many user groups have expressed concerns about the effects of  differential privacy on redistricting.  They largely but not exclusively concern two issues. First, ``One Person, One Vote'' case law calls for balancing population across the electoral districts in a jurisdiction, whether small like city council districts or large like congressional districts.
Most states balance  congressional districts to within one person based on Census counts.
Second, the most reliable legal tool against gerrymandering has been the Voting Rights Act of 1965 (VRA), which requires a demonstration of racially polarized voting (RPV). This RPV analysis is typically performed by statistical techniques that infer voting by race from precinct-level returns.  Many voting rights advocates worry that noising of Census data will confuse population balancing practices, and others worry that it will attenuate RPV signals, making it harder to press valid claims.

The Census Bureau has been commendably transparent about the development of \TopDown, making working code publicly available along with documentation and research papers describing the algorithm.
The complexity of the algorithm makes it extremely difficult to study analytically, so many people have sought to run it on realistic data.
However, since person-level Census data remain confidential for 72 years after collection, detailed input data for \TopDown is not public. Data users who would like to understand its impacts are left with two options: decades-old data or a limited demonstration data product.

In this paper, we get around the empirical obstacle by use of reconstructed block-level 2010 microdata for the state of Texas, and we try to understand the algorithm through theoretical analysis of a much-simplified toy algorithm, \Toy, that retains the two-stage, top-down structure of \TopDown but is much easier to analyze symbolically.
We investigate three questions about the count discrepancies created by \TopDown in units of census geography and ``off-spine'' aggregations like districts and precincts.

\vspace{-10pt}

\subparagraph{Hierarchical budget allocation.}
We derive easy-to-evaluate expressions for \Toy errors as a function of the privacy budget allocation. Error at higher levels of the geographic hierarchy impacts  lower-level counts with a significant discount, suggesting that bottom-heavy allocations may be optimal for accuracy on small geographies.
This is consistent with the small-district errors in our experiments with \TopDown.  For larger districts, a tract-heavy allocation gives greatest accuracy.  Equal allocation over the levels is a strong performer in both cases, making this a good choice from the point of view of multi-scale redistricting.

\vspace{-10pt}

\subparagraph{District construction.}
From there, we create further tests to study the impacts of district design. We compare hierarchically greedy  to geometrically greedy district-generation schemes, where the former attempt to keep large units whole and the latter attempt to build districts with short boundaries.   We find that the \Toy model gives errors very closely keyed to the fragmentation of the hierarchy, but that spatial factors damp out the primary role of fragmentation in the shift to the  \TopDown setting.

\vspace{-10pt}

\subparagraph{Robustness of linear regression.}
Finally, we consider the unweighted linear regressions commonly used to assess racial polarization in voting rights cases.  We find that the noise from both \Toy and \TopDown introduces an attenuation bias that seems alarming at first. However, unweighted linear regression on precincts is already vulnerable to major skews imposed by the inclusion of very small precincts.  For any reasonable way of counteracting that---trimming out the tiny precincts or weighting the regression by the number of votes cast---the instability introduced by \Toy and \TopDown all but vanishes.

\medskip

Our investigation is set up to answer questions about the status quo workflow in redistricting.
As usual with studies of differential privacy, a finding that DP unsettles the current practices might lead us to call to refine the way it is applied, but might equally lead us to interrogate the traditional practices and seek next-generation methods for redistricting.
In particular, it is clear that the practice of {\em one-person} population deviation across districts was never reasonably justified by the accuracy of Census data nor required by law, and the adoption of differential privacy might give redistricters occasion to reconsider that practice. We make a similar observation about the way that racially polarized voting analysis is commonly performed in expert reports.
On the other hand, by focusing on decisions still to be announced like the privacy budget and its allocation over the hierarchy, we are able to make recommendations that can assist the Bureau in protecting privacy while attending to the important concerns of user groups.

\section{Background on Census and redistricting}

\subsection{The structure of Census data and the redistricting data products}

Every ten years the U.S. Census Bureau attempts a comprehensive collection of person-level data---called {\em microdata}---from every household in the country.
The microdata are  confidential, and are only published in aggregated tables subject to disclosure avoidance controls.
The Decennial Census records information on the sex, age, race, and ethnicity for each member of each household, using categories set by the
Office of Management and Budget \cite{Census-sum}.  The 2020 Census used six primary racial categories: White, Black, American Indian, Asian, Native Hawaiian/Pacific Islander, and Some Other Race.  An individual can select these in any combination but must choose at least one, creating $2^6-1=63$ possible choices of race.  Separately,
{\em ethnicity} is represented as a binary choice of Hispanic/Latino or not.

The 2010 Census divided the nation into over 11 million small units called {\em census blocks} which nest in larger geographies in a six-level ``central spine'':  nation---state---county---tract---block group---block. %
Counts of different types are provided with respect to these geographies.
This tabular data is then used in an enormous range of official capacities, from the apportionment of seats in the U.S. House of Representatives to the allocation of many streams of federal and state funding.
The redistricting (PL 94-171) data includes four such tables: H1, a table of housing units whose types are occupied/vacant; and four tables of population, P1 (63 races), P2 (Hispanic, and 63 races of non-Hispanic population), and P3/P4 (same as P1/P2 but for voting age population).
Each table can be thought of as a \emph{histogram}, with each included type constituting one histogram \emph{bin}.
For instance, in table P1 there is 1 person in the $t=$White+Asian bin in the Middlesex County, MA, block numbered 31021002.

Treating the 2010 tables as accurate, it is easy to infer information not explicitly presented in the tables. For instance, the same bin in the P3 table (race for voting age population) also has a count of 1, implying that there are no White+Asian people under 18 years old in block 31021002.  This is the beginning of a {\em reconstruction} process that would enable an attacker, in principle, to learn much of the person-level microdata behind the aggregate releases.

\subsection{Disclosure avoidance}

Title 13 of the U.S. Code requires the Bureau to take measures to protect the
privacy of respondents' data \cite{Title13}. In the 2010 Census, this was largely achieved by an ad hoc mechanism called {\em data swapping}: a Bureau employee manually swapped data between small census blocks to thwart
re-identification.
In 2020, swapping is no longer considered adequate to protect against more sophisticated (but mathematically straightforward) data attacks that seek to reconstruct the individual microdata.
An internal Census Bureau study concluded that data swapping was unacceptably vulnerable: Census staff were able to reconstruct the 2010 Census responses of---and correctly reidentify---tens of millions of people.

With the reconstruction/reidentification threat in mind, the Bureau
has developed an algorithm called \TopDown \cite{TopDown}, which begins with a noising step that is
{\em differentially private}, following a mathematical formalism that provides rigorous guarantees against information disclosure \cite{DMNS}.
Differentially private algorithms obey a quantifiable limit to how much the output can depend on an individual record in the input.
The relationship of output to input is specified by a tuneable parameter, $\eps$, often called the {\em privacy budget}.
When $\eps\to\infty$, the output approaches equality to the input (high risk of disclosure).  When $\eps\to 0$, the output bears no resemblance to the input whatsoever (no risk of disclosure).
Like a fiscal budget, the privacy budget can be allocated until it is fully spent, in this case by spending parts of the budget on particular queries and on levels of the hierarchy.

\TopDown takes an individual-level table of census data and creates a `synthetic' dataset that will be used in its place to generate the PL 94-171 tables.
It can be thought of as taking as input a histogram with a bin for each person type (i.e., a combination of race, sex, ethnicity, etc.) and outputting an altered version of the same histogram. It proceeds in two stages. First, it privatizes the input histogram counts: it adds enough random noise to get the required level of differential privacy (according to the budget $\eps$). At this stage, it also allocates a portion of the total privacy budget for generating additional noisy histograms of data of particular importance to the Census Bureau. Second, \TopDown does post-processing on the noisy histograms to satisfy a handful of additional plausibility constraints. Among other things, post-processing ensures that the resulting histograms contain only non-negative integers, are self-consistent, and agree with the raw input data on a handful of \emph{invariants} (e.g., total state population).

The overall privacy guarantees of \TopDown are poorly understood.
In this paper, we design a simpler cousin of \TopDown nicknamed \Toy and we explore the properties of both \Toy and \TopDown, primarily focusing on reconstructed Texas data from 2010.

\subsection{The use of Census products for redistricting}
The PL 94-171 tables are the authoritative source of data for the purposes of apportionment to the U.S. House of Representatives, and with a very small number of exceptions also for within-state legislative apportionment.
The most famous use of population counts is to decide how many members of the 435-seat House of Representatives are assigned to each state.
In ``One person, one vote'' jurisprudence initiated in the {\em Reynolds v. Sims} case of 1964, balancing Census population is required not only for Congressional districts within a state but also for districts that elect to a state legislature, a county commission, a city council or school board, and so on \cite{ReynoldsVSims, WesberryVSanders, AveryVMidlandCounty}.

Today, the Congressional districts within a state usually balance total population extremely tightly:  each of Alabama's seven Congressional districts drawn after the 2010 Census has a total population of either
682,819 or 682,820 according to official definitions of districts and the Table P1 count, while Massachusetts districts all have a population of 727,514 or 727,515. %
Astonishingly, though no official rule demands it, more than half of the states maintain this ``zero-balancing'' practice (no more than one person deviation) for Congressional districts \cite{NCSL}.
This ingrained habit of zero-balancing districts to protect from the possibility of a malapportionment challenge is the first source of worry in the redistricting sphere.
If disclosure avoidance practices introduce some systematic bias---say by creating significant net redistribution towards rural and away from urban areas---then it becomes hard to control overall malapportionment, which could in principle trigger constitutional scrutiny.  In the end, redistricters may not care very much how many people live in a single census block, but it could be quite important to have good accuracy at the level of a district.

The second major locus of concern for redistricting practitioners is the enforcement of the Voting Rights Act (VRA).
Here, histogram data is used to estimate the share of voting age population held by members of minority racial and ethnic groups. Voting rights attorneys must start by satisfying three threshold tests without which no suit can go forward.
\begin{itemize}
\item \textbf{Gingles 1}:  the first ``Gingles factor'' in VRA liability is satisfied by creating a demonstration district  where the minority group makes up over 50\% of the voting age population.
\item \textbf{Gingles 2-3}: the voting patterns in the disputed area must display {\em racial polarization}.
The minority population is shown to be cohesive in its candidates of choice, and bloc voting by the majority prevents these candidates from being elected.  In practice, inference techniques like linear regression or so-called ``ecological inference'' are used to estimate voting preferences by race.
\end{itemize}

Since the VRA has been a powerful tool against gerrymandering for over 50 years, many worry that
even where
the raw data would clear the Gingles preconditions,
the noised data will tend towards uniformity---blocking deserving plaintiffs from a cause of action.

\section{Census \TopDown and \Toy}

\subsection{Setup and notation}
For the Census application, the data universe is a set of {\em types}: for instance, the redistricting data (the PL 94-171) has the types
$T=T_R\times T_E \times T_{VA} \times T_H$, where $T_R$ is the set of 63 races, $T_E$ is binary for ethnicity (Hispanic or not), $T_A$ is binary for age (voting age or not), and $T_H$ is the set of housing types.  (The fuller decennial Census data has more types.)%

A \emph{hierarchy} $H$ is a rooted tree of some depth $d$, so that every leaf has distance $\le d-1$ from the root.  We will usually assume the hierarchy has uniform depth, so that every leaf is exactly $d-1$ away from the root.
For node $h\in H$, let $n(h)\in \bbN$ be the number of children of $h$ in the tree, and let $\ell(h)$ be the level of node $h$.
A hierarchy is called \emph{homogeneous} if each node at level $\ell$  has the same number  of children, denoted $n_\ell$.
Let $H_\ell$ denote the set of nodes at level $\ell$, so that the set of leaves is $H_d$ in the uniform-depth case.
Label the root of the tree $h=1$.
We adopt an indexing of the tree and refer to the $i$th child of $h$ as $h_i$;
 the parent of any non-root node $h$ is denoted $\hat h$.
In Census data, the hierarchy represents the large and complicated set of nested geographical units, from the nation at the root down to the census blocks at the leaves.  The standard hierarchy has the six levels (nation---state---county---tract---block group---block) described above.

We associate with hierarchy $H$ and types $T$ a set of
{\em counts} $A_{H,T} = \{a_{h,t} \in \bbN\}_{h\in H, t\in T}$, where $a_{h,t}$ is the population of type $t$ in unit $h$ of census geography. We say $A_{H,T}$ is \emph{hierarchically consistent} if the counts add up correctly:  for every non-leaf $h$ and every $t$, we require $a_{h,t} = \sum_{i\in [n(h)]} a_{h_i,t}$.
For a singleton $T$, we write $A_H = \{a_h\}$.
We set an {\em allocation} $(\eps_1,\dots,\eps_d)$ breaking down the privacy budget $\eps=\sum \eps_i$ to the different levels of the hierarchy.

Our {\em queries} will always be counting queries, so that for instance
$q_{F,44}(h)$ returns the number of 44-year-old females in geographic unit $h$.
This particular query is part of a ``sex by age'' {\em histogram} $Q_{sex,age} = \{q_{s,a} : s \in T_S, a \in T_A\}$, which partitions $T$ into {\em bins} by sex and age.  In this language, $q_{F,44}$ is a bin of the sex-by-age histogram.
By slight abuse of notation, we will use the same terminology for the queries and their outputs, so that the histogram can be thought of as the collection of queries or the collection of counts.
Similarly, the ``voting age by ethnicity by race'' histogram consists of a query for each combination of the $2\times2 \times 63$ possible combinations of the three attributes.

\subsection{\Toy and \TopDown}

The Bureau's \TopDown and our simplified \Toy are both algorithms for releasing privatized population counts for every $h\in H$.  That is, these algorithms protect privacy by noising the data histograms.
\TopDown releases not just total population counts, but counts by type.
We will define {\em single-attribute} and {\em multi-attribute} versions of \Toy that noise $A_H$ and $A_{H,T}$, respectively, where consistency must hold for each type $t$.

\TopDown and \Toy share the same two-stage structure. Starting with hierarchically consistent raw counts $a$, the \emph{noising stage} generates differentially private counts $\ahat$. The \emph{post-processing stage} solves a constrained optimization problem to find noisy counts $\alpha$ that are close to the $\ahat$ values while satisfying hierarchical consistency and other
requirements.
\TopDown is named after the iterative approach to post-processing:  one geographic level at a time, starting at the top (nation) and working down to the leaves (blocks).
We sketch the noising and post-processing here, and we describe them in Appendix~\ref{app:ToyTop-description} in more detail.

The simple \Toy model can be run in a single-attribute version (only counts $A_H$), a multi-attribute version (counts by type $A_{H,T}$), or in multi-attribute form enforcing non-negativity.
The single-attribute version is easy to describe:  level by level, random noise values are selected from a Laplace distribution with scale $1/\eps_\ell$ and added to each count, replacing each $a_h$ with $\ahat_h=a_h+L_h$.
Then, working from top to bottom, the noisy $\ahat_h$ are replaced with the closest possible real numbers $\alpha_h$ satisfying hierarchical consistency.
Multi-attribute \Toy is defined analogously, but using $A_{H,T}$ instead of $A_H$ and requiring hierarchical consistency within each type $t\in T$.
Non-negative \Toy adds the inequality requirement that  $\alpha_h\ge 0$.

\TopDown is structurally similar but much more complex, with more kinds of privatized counts in the noising stage and a great many more constraints in the post-processing stage, including integrality.
The privatized counts computed by \TopDown are specified by a collection of histograms (or complex queries) called a \emph{workload} $W$. %
For each bin of each histogram in the workload and for each node $h$ in the geographic hierarchy, \TopDown adds geometric noise to the count.
The post-processing step finds the closest integer point that satisfies the requirements given by hierarchical consistency, non-negativity, as well as additional conditions given as invariants and structural inequalities.  For example, any block with zero households in the raw counts must have zero households and zero population in the output adjusted counts. Together, the invariants, structural inequalities, integrality, and non-negativity make this optimization problem very hard. The problem is NP-hard in the worst case and \TopDown cannot always find a feasible solution.
There is a sophisticated secondary algorithm for finding approximate solutions that is beyond the scope of this paper.

\Toy is simple enough that solutions can often be obtained symbolically.
\Toy simplifies the noising stage by fixing the workload to be the detailed workload partition $Q_{detailed} = \left\{\{t\}\right\}_{t\in T}$ consisting of all singleton sets and using the continuous Laplace Mechanism instead of the discrete Geometric Mechanism. It simplifies the post-processing stage by dropping invariants, structural inequalities, integrality, and non-negativity.
When negative answers are permitted, multi-attribute \Toy is equivalent to executing $|T|$ independent instances of single-attribute \Toy on inputs $A_{H,t} = \{a_{h,t}\}_{h\in H}$ for each $t \in T$.
As a result, many of our analytical results for single-attribute \Toy extend straightforwardly to multi-attribute \Toy (allowing negative answers) by scaling by a factor of $|T|$ in appropriate places.

\section{Methods}

We use both analytical and empirical techniques in this work. This section describes our high-level empirical approach: what algorithms and raw data we used and how we used them. See Appendix~\ref{app:methods} for more details.
We repeatedly ran \TopDown and \Toy in various configurations on a reconstructed person-level Texas dataset created by applying a reconstruction technique to the block-level data from the 2010 Census, following \cite{Hawes} based on \cite{DinurNissim}.
The reconstructed microdata records---obtained from collaborators---contain block-level sex, age, ethnicity, and race information consistent with a collection of tables from 2010 Census Summary File 1.

We executed 16 runs of \TopDown with each of 20 different allocations of the privacy budget across the five lower levels of the national census geographic hierarchy: $\eps = \eps_{2}+\eps_{3}+\eps_{4}+\eps_{5}+\eps_{6}$.
The 20 allocations consist of five different splits across the levels (Table~\ref{tab:budget-splits}) for each of four total budgets $\eps \in \{0.25, 0.5, 1.0, 2.0\}$.
\TopDown operates on the six-level Census hierarchy and requires specifying $\eps_{1}$. In our experiments, we ran \TopDown with a fixed total privacy budget $\eps_{total} = 10$, with $\eps_{1} = 10 - \eps$. Because the nation-level budget is so much higher than the lower level budgets, we omit further discussion of it.
The \TopDown workload was modeled after the workload used in the 2018 End-to-End test release, omitting household invariants and queries.

We also ran three variants of \Toy (single-attribute, multi-attribute, and non-negative) on a simplified version of the same data 2010 data. We executed 16 runs of each variant with each of five different splits of the privacy budget across the five lower levels of the census geographic hierarchy (Table~\ref{tab:budget-splits}), fixing the total budget for those five levels at $\eps = 1$.
The data was derived from the reconstructed Texas data simplified to include only seven distinct types: one for the total Hispanic population and one for each of six subgroups of the non-Hispanic population based on race (White; Black; American Indian; Asian; Native Hawaiian/Pacific Islander; and Some Other Race or multiple races).
Post-processing for single-attribute \Toy was implemented in NumPy, while post-processing for multi-attribute and  non-negative \Toy used a {\tt Gurobi} solver.

\section{Hierarchical budget allocation}\label{sec:allocation}

The relationship of the hierarchical allocation $(\eps_1,\dots,\eps_d)$ to various measures of output accuracy is not obvious.  On one hand, it might seem that higher values of $\eps_d$ (the block-level budget) will best promote accuracy at the block level, for a fixed $\eps$.  But on the other hand, imposing hierarchical consistency forces lower levels to be consistent with the totals at higher levels, which means that noise at higher levels can trickle down to lower levels.  These  competing effects  create tradeoffs that are hard to balance without further analysis.

\begin{table} \begin{minipage}{.52\linewidth}

\centering
{\footnotesize
\begin{tabular}{r|c|c|c|c|c}
& state& county & tract & BG & block\\
Split name & $\eps_{2}$ &$\eps_{3}$ &$\eps_{4}$ &$\eps_{5}$ &$\eps_{6}$ \\
\hline
equal & 0.2 & 0.2 & 0.2 & 0.2 & 0.2\\
state-heavy & 0.5 & 0.25 & 0.083 & 0.083 & 0.083 \\
tract-heavy &  0.083 & 0.167 & 0.5 & 0.167 & 0.083 \\
BG-heavy & 0.083 & 0.083 & 0.167 & 0.5 & 0.167 \\
block-heavy &  0.083  & 0.083  & 0.083 & 0.25  & 0.5
\end{tabular}}

\vspace{10pt}

\caption{Names of designated budget splits used in  \Toy and \TopDown runs below, each with a budget of $\eps_1=9$ on the nation and a total of 1 allocated below the national level.}
\label{tab:budget-splits}
\end{minipage}\hfill
\begin{minipage}{.38\linewidth}

\centering
\begin{tikzpicture}[scale=.67,sibling distance=6em,
  every node/.style = {shape=circle,
    draw, align=center}]

    \tikzstyle{level 1}=[sibling distance=30mm]
    \tikzstyle{level 2}=[sibling distance=10mm]

  \node {$\frac{7}{12}$}
    child { node {$\frac{1}{2}$}
        child { node {$0$} }
        child { node {$1$} }
    }
    child { node {$\frac{1}{4}$}
        child { node {\small $0$} }
        child { node {\small $0$} }
        child { node {\small $0$} }
        child { node {\small $1$} }
    }
    child { node {$1$ }
        child { node {$1$} }
        child { node {$1$} }
};
\end{tikzpicture}

\vspace{10pt}


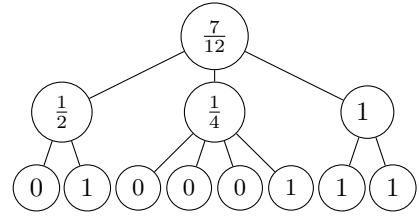
\captionof{figure}{A district in a three-level hierarchy. The $0/1$ weight of a leaf indicates its membership in the district; each non-leaf weight is the average of the node's children.}

\end{minipage}
\end{table}

\subsection{\Toy error expressions}
\label{sec:allocations-theory}

\begin{definition}[District, weights, error]
A \emph{district} $D \subseteq H_d$ is a subset of the leaves (blocks) of the hierarchy $H$.
For hierarchy $H$, a district $D$ induces \emph{weights}  $w_h \in [0,1]$ on the hierarchy nodes, defined recursively as follows:
\begin{itemize}
\item For each leaf $h \in H_d$, let $w_h = 1$ if $h\in D$ and $w_h=0$ otherwise.
\item For $\ell \le d-1$ and  $h\in H_\ell$, let  $w_h = \frac{1}{n(h)}\sdot \sum_{i \in [n(h)]} w_{h_i}$ be the average of the weights of the children.
\end{itemize}
\end{definition}

In a homogeneous hierarchy, we can observe that each $w_h$ equals the fraction of the leaves descended from $h$ that belong to $D$. In particular, the root weight is $w_1 = |D|/|H_d|=1/k$ if there are $k$ districts of equal population made from nodes of equal population.

For node $h\in H$, we record the {\em error} $E_h =\alpha_h - a_h$  introduced by $\Toy$ to the count $a_h$.
The total error over district $D$ is $E_D = \sum_{h \in D} E_h$. Let $\hat h$ denote the parent of node $h$.

\begin{restatable}[Error expressions]{theorem}{ErrorDistrictsTheorem}
\label{thm:error_district}
$E_1 = L_1$. For $\ell \in \{2,\dots,d\}$ and non-root node $h_i \in H_{\ell}$, and for every district $D$ with associated weights $w_h$ on the nodes,
\begin{equation}
E_{h_i} = L_{h_i} + \frac{1}{n(h)} \left(E_h - \sum_{j \in [n(h)]} L_{h_j}\right),
\quad \quad
E_D = w_1 L_1 + \sum_{h \in H \setminus \{1\}} (w_{h} - w_{\hat h})L_{h}.
\end{equation}
\end{restatable}

We make several observations. First, our intuition that error at higher levels trickles down to lower levels is correct, but this effect is rather weak. The error at a child $h_i$ is determined by the parent error $E_h$ discounted by the degree $n(h)$, the number of siblings. This suggests that placing more budget at level $\ell$ is an efficient way to secure accuracy at that level, until a fairly extreme level of error at higher levels overwhelms the degree-based ``discount.''

Second, because the $L_h$ are all independent random variables with $\E(L_h) = 0$ and $\Var(L_h) = 8/\eps_{\ell(h)}^2$, the theorem provides the following expression for variance that we use repeatedly.

\begin{corollary}[Error expectation and variance]
\label{cor:error-variance}
For all $D\subseteq H_d$ and associated weights $w_h$, the expected error and error variance produced by \Toy satisfy
$\E(E_D) = 0$ and
\begin{equation}
\label{eqn:var_district}
    \Var(E_D) = \frac{8w_1^2}{\eps_1^2} + \sum_{\ell = 2}^d \left(\frac{8}{\eps_\ell^2} \sdot \sum_{h \in H_\ell}(w_{h} - w_{\hat h})^2\right).
\end{equation}
\end{corollary}

Third, we get a more explicit expression if restricting to homogeneous hierarchies $H$.
Consider the case of a singleton district $\{h\}$ made of a single census block $h\in H_d$.

\begin{corollary}[Error variance, homogeneous case]
\label{cor:error-homogeneous}
The \Toy error for a single block $h\in H_d$ satisfies
\begin{equation}
    \Var(E_h) = \frac{8}{\eps_1^2 (n_1 \cdots n_{d-1})^2}
        +
        \sum_{\ell=2}^d \frac{8n_{\ell-1}(n_{\ell-1} - 1)}{\eps_\ell^2 (n_{\ell-1}  \cdots n_{d-1})^2}.
\end{equation}
\end{corollary}

\begin{figure}[ht]
\begin{minipage}{.53\linewidth}
\centering
\includegraphics[width=\textwidth]{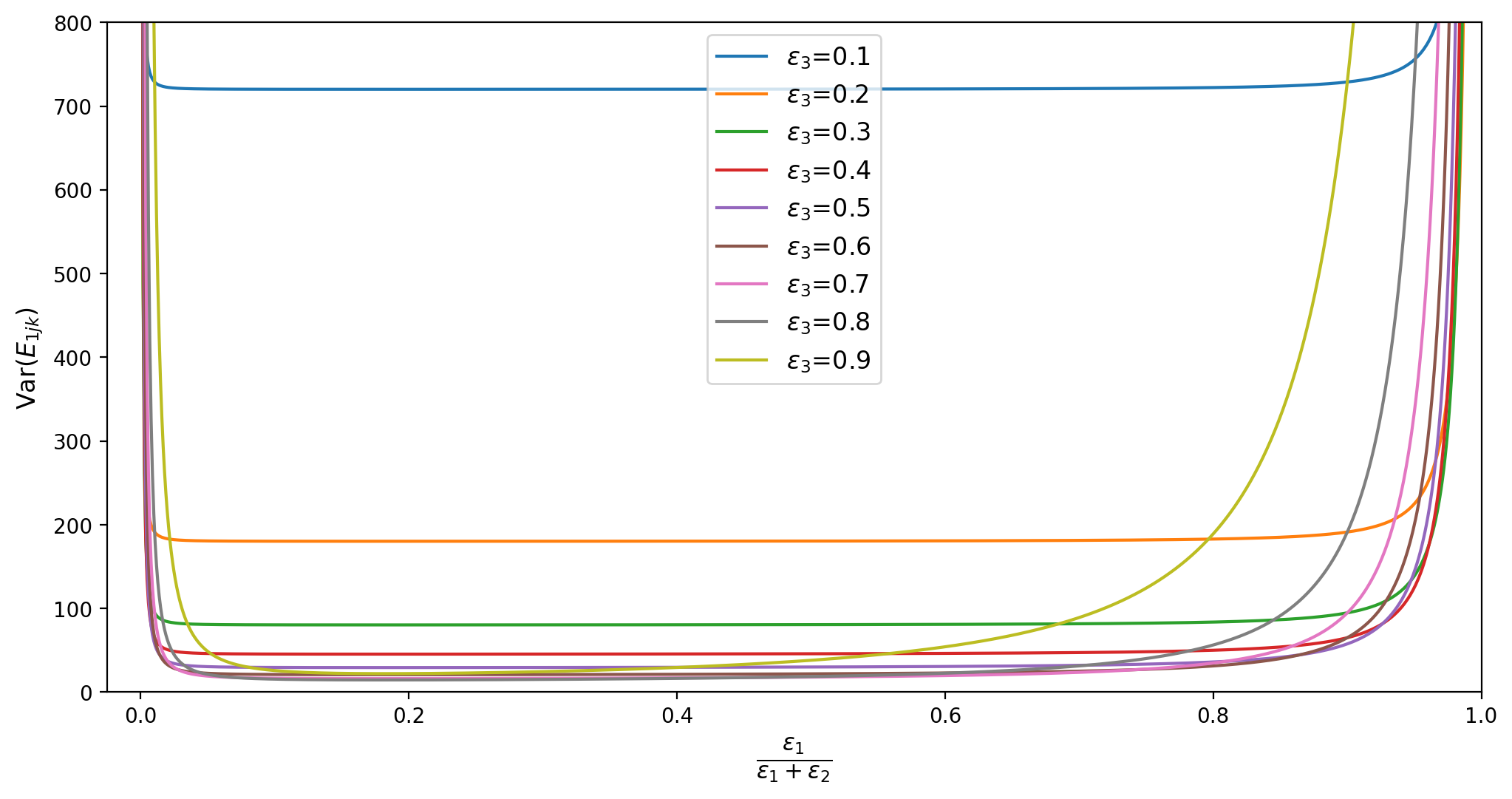}

\caption{\Toy error variance for a leaf node in the three-level hierarchy with $n_1=n_2=10$ and $\eps=1$.  The curves show varying $\eps_3$ (colors) and the relative balance of $\eps_1$ and $\eps_2$ ($x$-axis).
\label{fig:toydown-variance}}
\end{minipage}
\hfill
\begin{minipage}{.45\linewidth}

\centering

\vspace{20pt}

{\small \begin{tabular}{c|c|c}
 $\eps$ & Allocation & $L^1$ error\\
\hline
1.0 & (.16,\ .16,\ .16,\ .16,\ .16,\ .2) & 0.03 \\
1.0 & (.2,\ .16,\ .16,\ .16,\ .16,\ .16) & 0.03 \\
1.0 & (.1,\ .1,\ .1,\ .1,\ .1,\ .5) & 0.02 \\
1.0 & (.02,\ .02,\ .02,\ .02,\ .02,\ .9) & 0.03 \\
1.0 & (.66,\ .30,\ .01,\ .01,\ .01,\ .01) & 0.09
\end{tabular}}
\vspace{40pt}
\captionof{table}{$L^1$ error measurements from selected \TopDown runs on reconstructed Texas data. The allocation $(\eps_1,\dots,\eps_6)$ goes from the nation $\ell=1$ down to census blocks at $\ell=6$.}\label{topdown-table}\end{minipage}
\end{figure}

Figure~\ref{fig:toydown-variance} plots this expression for various ways of splitting a total privacy budget of $\eps = 1$ across a three-level hierarchy with $n_1 = n_2 = 10$.
The minimum of $f(x_1,\dots,x_d) = \sum_{\ell=1}^d {a_\ell}/{x_\ell^2}$ subject to $\sum_\ell x_\ell = \eps$ and  $x_\ell \ge 0$ is achieved at
$
x_\ell = {\eps a_\ell^{1/3}}/{\sum_{i} a_i^{1/3}}$ for all $\ell$.
For the example in Figure~\ref{fig:toydown-variance}, the minimum-variance split is $(\eps_1, \eps_2, \eps_3) = (0.038, 0.171, 0.791)$ with variance $14.52$.
(See accompanying \href{https://colab.research.google.com/drive/1qal-xj2SDt2zOYT4VAcVnew25zUwkA4I?usp=sharing}{CoLab notebook}.)
One important note in interpreting Figure~\ref{fig:toydown-variance}
is that these variance numbers are absolute and don't depend on knowing population counts for the nodes of the hierarchy.  They are simply based on sampling Laplace noise with the given parameters.  If a variance of about 15 in the bottom-level counts is too high to be tolerated in an application, one would have to increase $\eps$ to achieve lower variance.

\vspace{-20pt}
\subsection{Empirical error experiments in \TopDown}
\label{sec:allocations:experiments}

Next, we move to \TopDown, which requires the use of input data.
First, using reconstructed 2010 Texas data, we varied the relative allocation vector and the total $\eps$, then measured the effects with an $L^1$ error metric included in the Census code \cite{End-to-end-L1}.
This is a measure of block-level error: it adds the magnitudes of changes in the bins, then divides by twice the total population in the histogram. %

Table~\ref{topdown-table} reports a small selection of the 100+ different scenarios explored.
In general, the lowest error outcomes were observed in a few scenarios:  when the budget was distributed near-equally to the levels of the hierarchy, and when half of the available budget was placed at the bottom level---beyond $\eps_d=\eps/2$, further bottom-weighting gave diminishing returns in block-level accuracy.

But a budget allocation that produces small block-level errors may not produce small errors for {\em districts}, depending  on  the  degree  of  cancellation  or  correlation.   Next,  we  use random district generation to understand the effects of off-spine aggregation.
In particular, we employ the Markov chain sampling algorithm called {\em recombination} (or \ReCom), which runs an elementary move that fuses two neighboring districts and re-partitions the double-district by a random balanced cut to a random spanning tree \cite{ReCom}.

\begin{figure}[ht]
\centering
\begin{tikzpicture}[scale=.7]
\node at (0,0) {\includegraphics[width=1.4in]{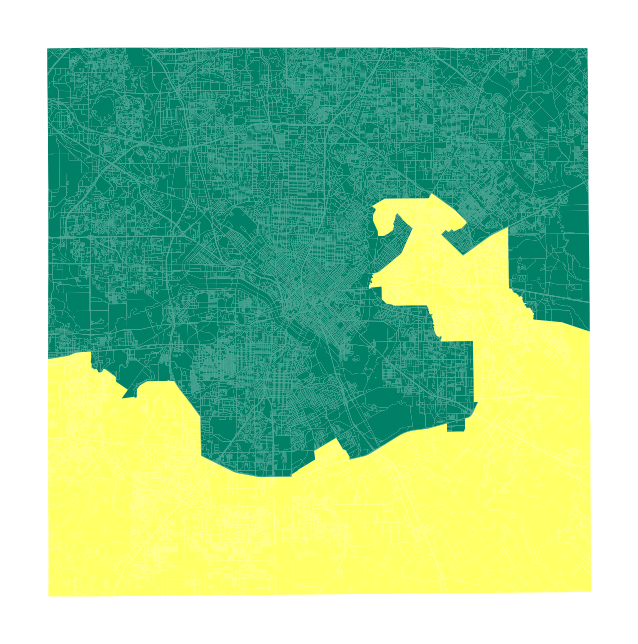}};
\node at (6.5,0) {\includegraphics[width=1.4in]{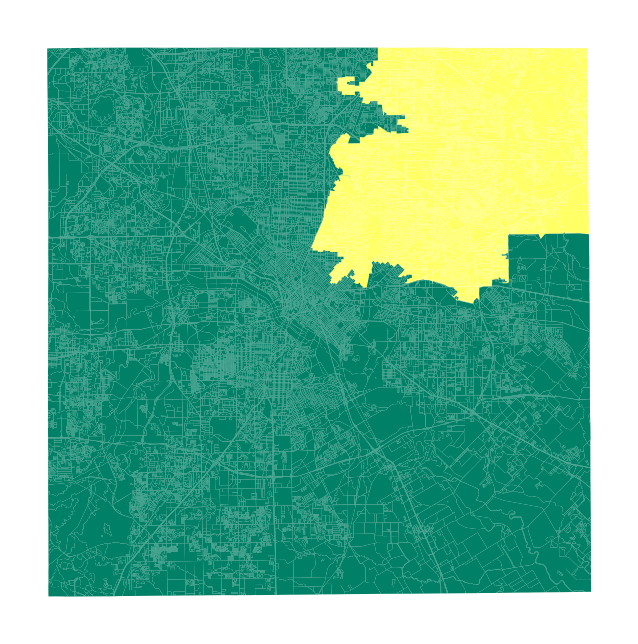}};
\node at (13,0) {\includegraphics[width=1.4in]{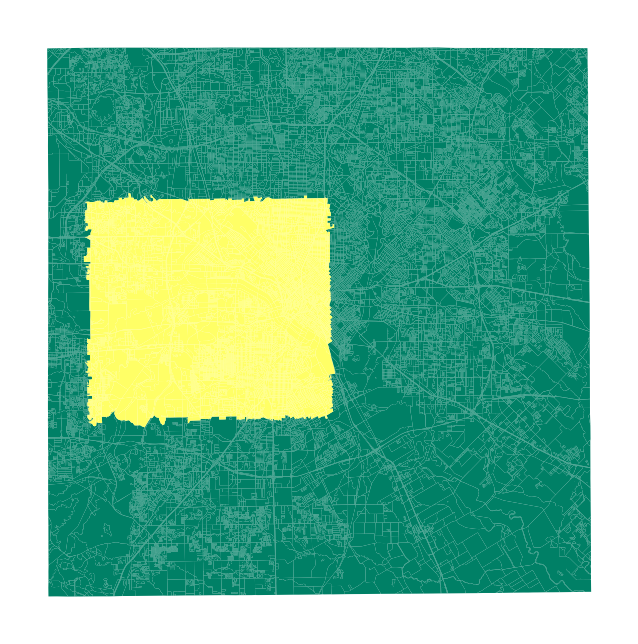}};
\end{tikzpicture}
\caption{Three sample districts (yellow) in Dallas County, each within two percent of the ideal population for $k=4$ districts.  These are drawn by tract \ReCom, block \ReCom, and a square-favoring algorithm, respectively.}
\end{figure}

\begin{figure}[ht]
    \centering
\begin{tikzpicture}
\node at (0,0)    {\includegraphics[width=6.4cm]{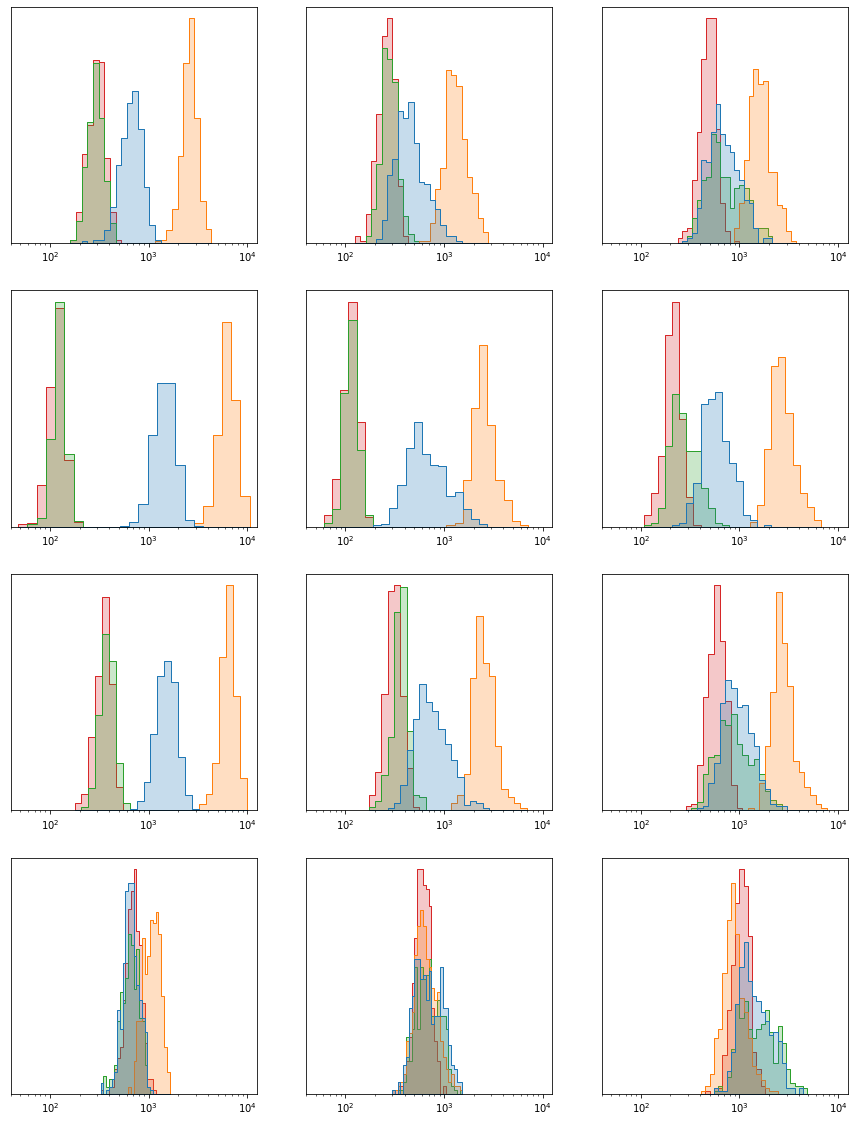}};
\node at (0,5.4) {$k=4$ districts};
\node at (-2.2,4.7) {\footnotesize \Toy};
\node at (0,4.8) {\footnotesize \Toy};
\node at (0,4.5) {\scriptsize non-neg};
\node at (2.15,4.7) {\footnotesize \TopDown};
\node at (-3.45,3.25) [rotate=90] {\footnotesize equal};
\node at (-3.45,1.2) [rotate=90] {\footnotesize tract-heavy};
\node at (-3.45,-.8) [rotate=90] {\footnotesize BG-heavy};
\node at (-3.45,-3.1) [rotate=90] {\footnotesize block-heavy};

\begin{scope}[xshift=7cm]
\node at (0,0) {\includegraphics[width=6.5cm]{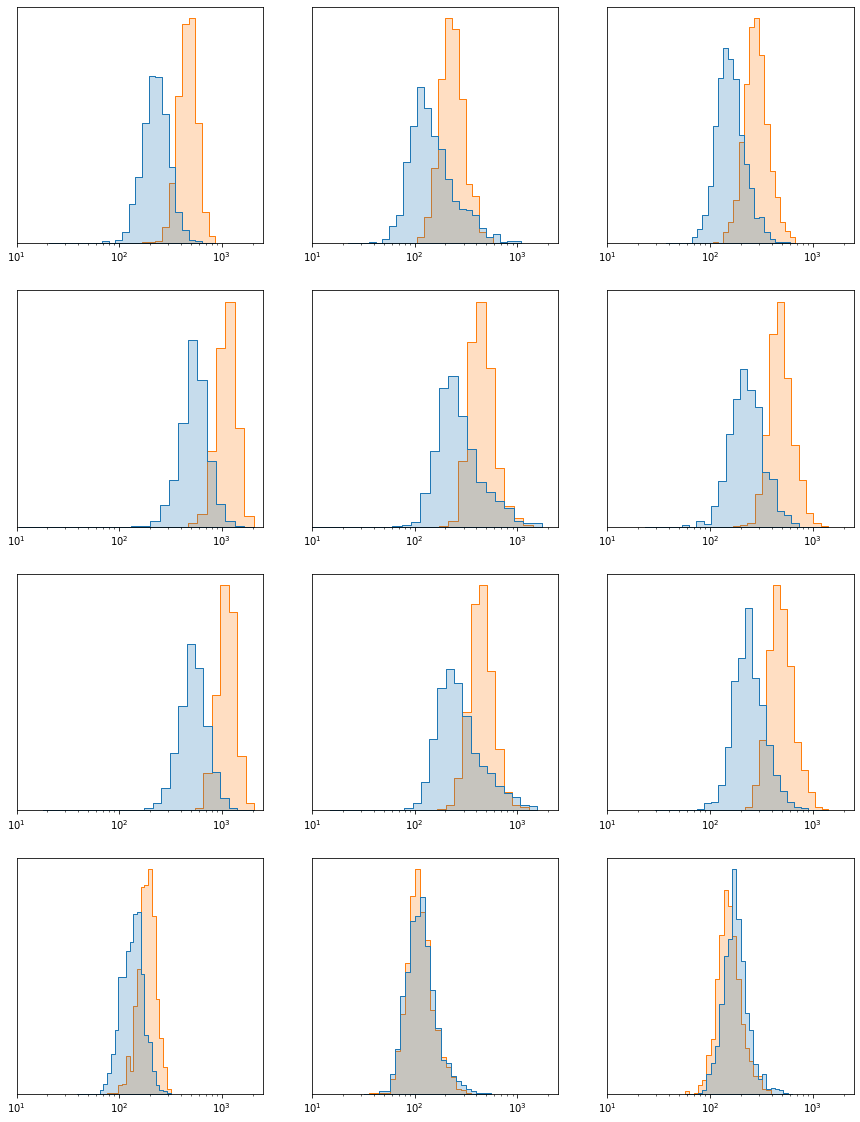}};
\node at (0,5.4) {$k=175$ districts};
\node at (-2.2,4.7) {\footnotesize \Toy};
\node at (0,4.8) {\footnotesize \Toy};
\node at (0,4.5) {\scriptsize non-neg};
\node at (2.15,4.7) {\footnotesize \TopDown};
\end{scope}

\node at (3.5,-4.7) {Green: tract \ReCom, \quad Red: tract \Disconn, \quad Blue: block \ReCom,\quad  Orange: block \Disconn
};

\end{tikzpicture}
\caption{These histograms
show district-level error on a log scale for various
combinations of budget splits (rows), district-drawing algorithms (colors), and noising algorithms (columns).  We include both large districts and small districts, dividing the county into $k=4$ and $k=175$ equal parts.
Each histogram displays 400 values, one for each district drawn by the specified  algorithm, plotting the mean observed district-level population error magnitude over 16 executions of the noising algorithm using the specified budget allocation.}
    \label{fig:mean_errormag_with_log_axis}
\end{figure}

We begin with county commission districts in Dallas County, where $k=4$.  Since the 2010 population of Dallas County was roughly 2.4 million, each district will have roughly 600,000 people, making them nearly as big as congressional districts and much larger than tracts.  We also include divisions of the county into $k=175$ districts of between 13,000 and 14,000 people each for a small-district comparison.
Figure~\ref{fig:mean_errormag_with_log_axis} plots the data from our experiments on a logarithmic scale. Each histogram displays 400 values, one for each district drawn by the specified district-drawing algorithm; each value is the mean observed district-level population error magnitude over 16 executions of the specified hierarchical noising algorithm using the specified budget allocation.

First, consider two unrealistic forms of district-generation:  tract \Disconn (red) and block \Disconn (orange), which randomly choose units of the specified type until assembling a collection with the appropriate population.  These are unrealistic because they do not form connected districts; here, they are used to illustrate the effects of aggregation, neglecting spatial factors entirely.
We see in Figure~\ref{fig:mean_errormag_with_log_axis}
that block-based methods generate hugely more error than tract-based methods, except if the budget allocation is concentrated at the bottom of the hierarchy.
The effect is stronger for \Toy (in keeping with Theorem~\ref{thm:error_district}), but is easily observed for \TopDown as well.

We compare that with the more realistic district-generation algorithm   block \ReCom (blue), which builds compact and connected districts out of block units.  This tends to give error levels in between the extremes set by the other two.  Likewise, tract \ReCom (green) builds compact and connected districts from tracts.
One reasonable mechanism by which  \ReCom has much lower error than  \Disconn is that \ReCom districts will tend to have  higher ``hierarchical integrity,'' keeping higher-level units whole just by virtue of being connected and plump.  The interior of \ReCom districts will thus contain many whole block groups and tracts.  Near the boundary, block groups and tracts are more fragmented, leaving the corresponding block-level errors uncancelled. These fragmentation ideas are explored more fully in Section~\ref{sec:fragmentation-definition} and some sample districts are depicted here.

The cancellation effect is significant: in most experiments, the error level for \ReCom districts is much closer to that of tract \Disconn than block \Disconn (recall the data is plotted on a logarithmic scale).
Overall, drawing districts out of larger pieces (e.g., using tract \Disconn instead of \ReCom, or \ReCom instead of block \Disconn) lowers error magnitude significantly in the best case and has little or no effect in the worst case.

Although tract \ReCom and tract \Disconn behave very similarly under \Toy, the compact districts perform noticeably worse than their disconnected relatives once we pass to the full complexity of \TopDown.  At first this seems puzzling, because compact and connected districts are being punished by the geography-aware \TopDown.  But the reason for this is apparent on further reflection: {\em spatial autocorrelation} is causing the post-processing corrections to move nearby tracts in the same direction, impeding the cancellation that makes counts usually more accurate on larger geographies.

In the end, the story that emerges from these investigations is that, with full \TopDown, the best accuracy that can be observed for large districts occurs when they are made from  whole tracts and the allocation is tract-heavy; an equal split is not much worse.
For districts with population around 13,000, $\eps=1$ noising creates errors in the low hundreds for compact, connected districts, with the best performance for block-heavy allocations.  Again, an equal split is not much worse, suggesting that this might be a good policy choice for accuracy in districts across many scales.

\section{Geometrically compact vs hierarchically greedy districts}
\label{sec:fragmentation-definition}

The analysis above suggests that the district-level error $E_D$ will depend not only on the randomness of the noising algorithms, but also on the geometry of $D$ and the structure of $H$.
This section studies the hypothesis that districts that disrespect the geographical hierarchy will tend to have higher error magnitude.
This section defines the \emph{fragmentation score}, relates a district's fragmentation score to its error variance under \Toy, and compares the fragmentation of two simple district-drawing algorithms on homogeneous hierarchies and simple geographies.
Ultimately, we find that the explanatory value of the fragmentation score decays as we move to more realistic deployment of \TopDown.
This discrepancy raises important questions for future study: Which of the many additional features of \TopDown attenuates the fragmentation--variance relationship?

We define a score intended to capture the contribution to $\Var(E_D)$ of the shape of the district with respect to the hierarchy. %
Recall that $\hat h$ denotes the parent of node $h$.
\begin{definition}[Fragmentation score]
For $D\subseteq H_d$, let $\displaystyle\frag(D) =\sum_{h\in H} (w_{h} - w_{\hat{h}})^2$.
\end{definition}
Because weights are in $[0,1]$, the score obeys $0\le \frag(D)<|H|$ for all districts, with higher scores indicating the presence of more units that are only partially included in $D$.

This fragmentation score is reverse-engineered from the expression for the variance of district-level population errors when using \Toy with privacy divided equally across levels of the hierarchy (Corollary~\ref{cor:error-variance}): namely, $\Var(E_D) =  \frac{8d^2}{\eps^2}\left(w_1^2+\frag(D) \right)$. %
When the district $D$ itself is a random variable sampled from some distribution, the expected fragmentation $\E(\frag(D))$ is similarly related to $\Var(E_D)$. Namely, using the law of total variation, when each level gets $\eps/d$ privacy budget: $$\Var(E_D) =\E\left(\Var(E_D | D) \right) + \Var\left( \E(E_D | D) \right) = \E(\Var(E_D | D)) = \frac{8d^2}{\eps^2}(\E(\frag(D)) + \E(w_1^2)).$$
When $\eps$ is allocated unequally across levels, as for the other splits in Table~\ref{tab:budget-splits}, the  simple analytical relationship between the fragmentation score and the error variance breaks down.

Observe that a hierarchy $H$ does not capture all of the geometry relevant to district drawing. In particular, $H$ does not directly encode any information about block adjacency, and therefore we can't detect from $H$ that a district is contiguous. For algorithms to generate contiguous districts, we need to make use of the plane geometry associated to $H$.
We restrict our attention to the simplest case: homogeneous hierarchies (where every node on level $\ell< d$ has exactly $n_\ell$ children) and \emph{square tilings}. (where each unit on level $\ell$ is a square and has  $n_\ell$ children that cover it with a $\sqrt{n_\ell} \times \sqrt{n_\ell}$ grid tiling).

We analyze the fragmentation score for two simple district-drawing algorithms (see Appendix~\ref{app:frag-empirical}).
The \Greedy algorithm builds a district from the largest subtrees possible, only subdividing a subtree when necessary. It takes as input $H$ and $k \in \bbN$ and returns a district of size $N=\lfloor|H_d|/k\rfloor$, assembled by starting with the largest available units at random and adding units that are adjacent in the labeling sequence without passing size $N$, then allowing one partial unit, and so on recursively at lower levels.
Observe that \Greedy depends only on the hierarchy $H$.
The \Square algorithm  takes as input a square, homogeneous hierarchy $H$ and $k \in \bbN$ such that the district size is a perfect square, $|D| =  |H_d|/k={s_d}^2$. It outputs a uniformly random $s_d\times s_d$ square of blocks.

\begin{restatable}{theorem}{FragRandomTheorem}
\label{thm:random}
Let $D_{G}\sim \Greedy(H,k)$, $D_{\square}\sim \Square(H,k)$.
For  $n_1\cdot n_{2}\cdots n_{d-2} \ge k\ge 2$,
let $L = \argmin \{\ell : n_1\cdot n_2 \cdots n_{\ell} \ge k\}$.
\begin{equation*}
\E(\frag(D_{G}))
\le
\frac{k-1}{k^2}\sum_{\ell=1}^{L} n_\ell + \frac{1}{4}\sum_{\ell=L+1}^{d-1} n_\ell;
\quad
\E(\frag(D_{\square}))
\ge  \frac{2}{3}\left(\frac{\sqrt{n_1\dots n_{d-1}}}{\sqrt k} - \frac{11}2\right) \sqrt{n_{d-1}}.
\end{equation*}
\end{restatable}

Dallas County is nearly a perfect square shape, so it gives us an opportunity to set some roughly realistic parameters to evaluate these bounds.
There are 529 tracts in Dallas County, with an average of 3.2 blocks groups per tract and 26.4 blocks per block group, yielding 44,113 total blocks.
We can approximate these parameters by setting $d=4$, using $k=4$ as for the county commission districts, and setting $(n_1,n_2,n_3)=(484,4,25)$ which has a reasonably similar 48,400 blocks (as a result, $L=1$).
The bounds in the theorem say that
$\E(\frag(D_{G})) \le 98$ and $\E(\frag(D_{\square})) \ge 264$.
Note: for homogeneous hierarchies $H$ with equal-population leaves, the score $\frag(D_{G})$ is independent of algorithm randomness and can be computed exactly; for the above parameters $\frag(D_{G}) = 90.75$.
So the bound in the theorem is fairly tight, at least in this case.

To interpret the theorem, it is helpful to think of \Greedy as being hierarchically greedy and \Square as being geometrically greedy.  That is, the former is oriented toward using the biggest possible units and keeping them whole, so that spatial considerations are secondary;  the latter is oriented towards ``compact'' geographies with a lot of area relative to perimeter, and unit integrity is secondary.
The theorem shows that compactness alone (a function of the plane geometry) does not keep down the fragmentation score (a function of the hierarchy), and indeed the bounds get farther apart as the hierarchy gets larger and more complicated.
In Appendix~\ref{app:frag-empirical}, we compare these theoretical results to empirical district errors, finding that fragmentation tracks well with errors in \Toy, but that the complexity of the \TopDown model weakens the relationship, suggesting a need for more sophisticated tools.

\section{Ecological regression with noise}\label{sec:ER}

\subsection{Inference methods for Voting Rights Act enforcement}

When elections are conducted by secret ballot, it is fundamentally impossible to precisely determine voting patterns by race from the reported outcomes alone.  The standard methods for estimating these patterns use the cast votes at the precinct level, combined with the demographics by precinct, to infer racial polarization.  Because the general aggregate-to-individual inference problem is called ``ecological'' (cf. ecological paradox, ecological fallacy), the leading techniques are called {\em ecological regression} (ER) and {\em ecological inference} (EI).
It is rare that EI and ER do not substantively agree, and we focus on ER here because it lends itself to easily interpretable pictures.

ER is a simple linear regression, fitting a line to the data points determined by the precincts on a demographics-vs-votes plot.
A high slope (positive or negative) indicates a likely strong difference in voting preferences, which is necessary to demonstrate the Gingles 2-3 tests for a VRA lawsuit. %

\begin{figure}
    \centering
\begin{tikzpicture}[xscale=1.1]
\def\localwidth{4.2cm}

\node at (-4,8.5) {All precincts};
\node at (0,8.5) {Filtered precincts};
\node at (4,8.5) {Weighted precincts};

\node at (-4,6.5) {\includegraphics[width=\localwidth]{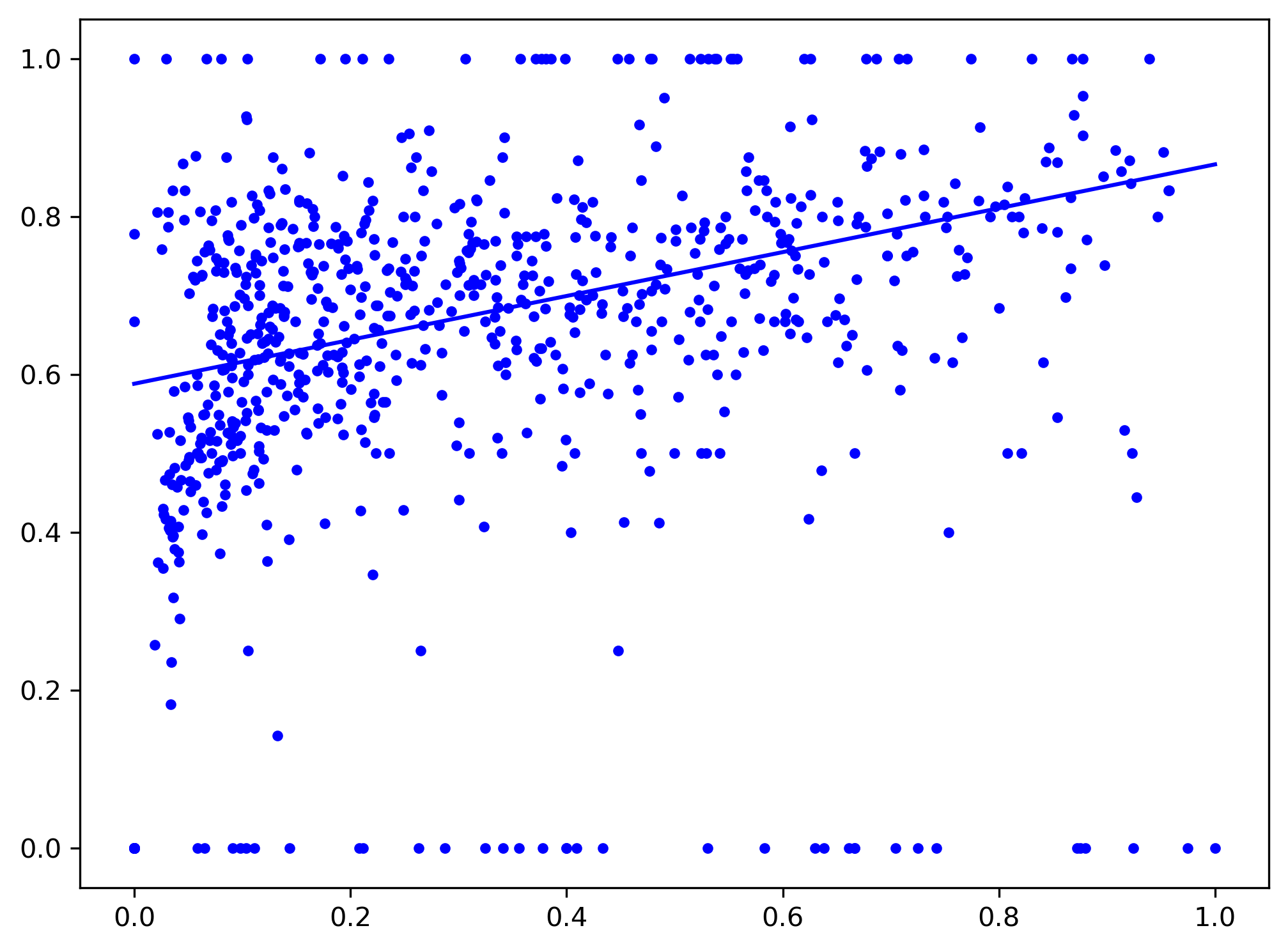}};

\node at (0,6.5) {\includegraphics[width=\localwidth]{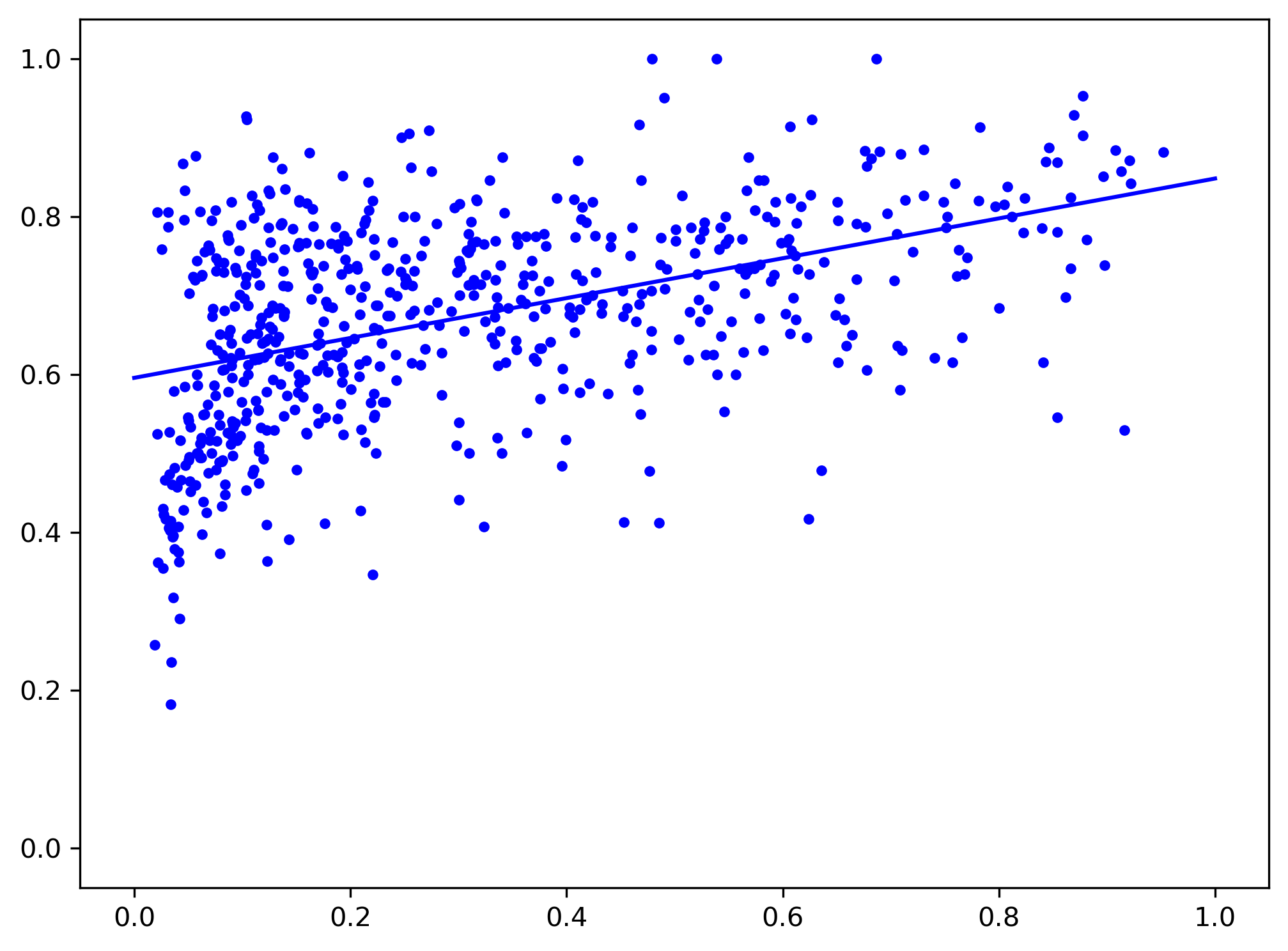}};

\node at (4,6.5) {\includegraphics[width=\localwidth]{figures/TopDown_Dallas_HVAP_weighted_er_just_blue.png}};

\node at (-4,3) {\includegraphics[width=\localwidth]{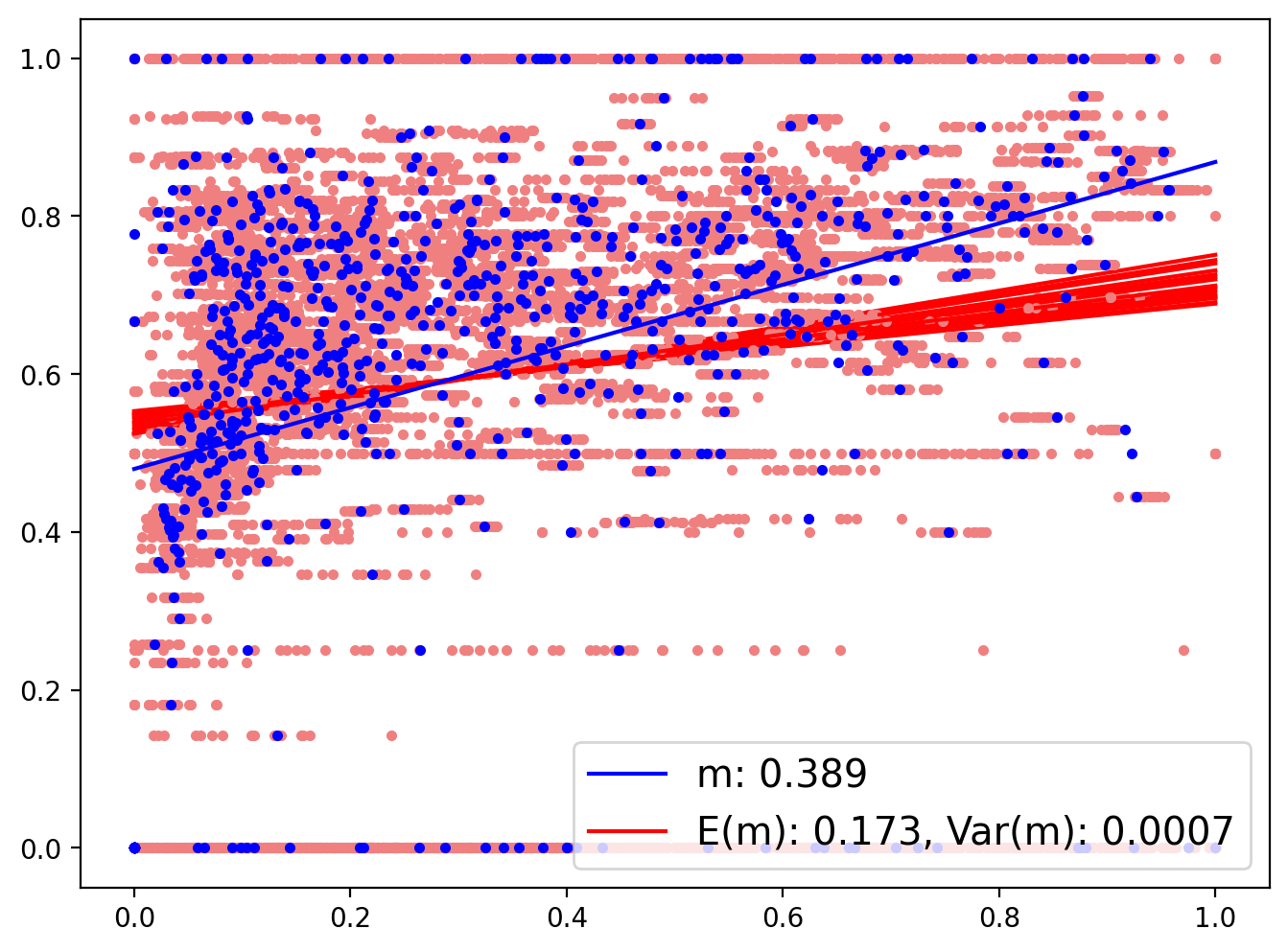}};

\node at (0,3) {\includegraphics[width=\localwidth]{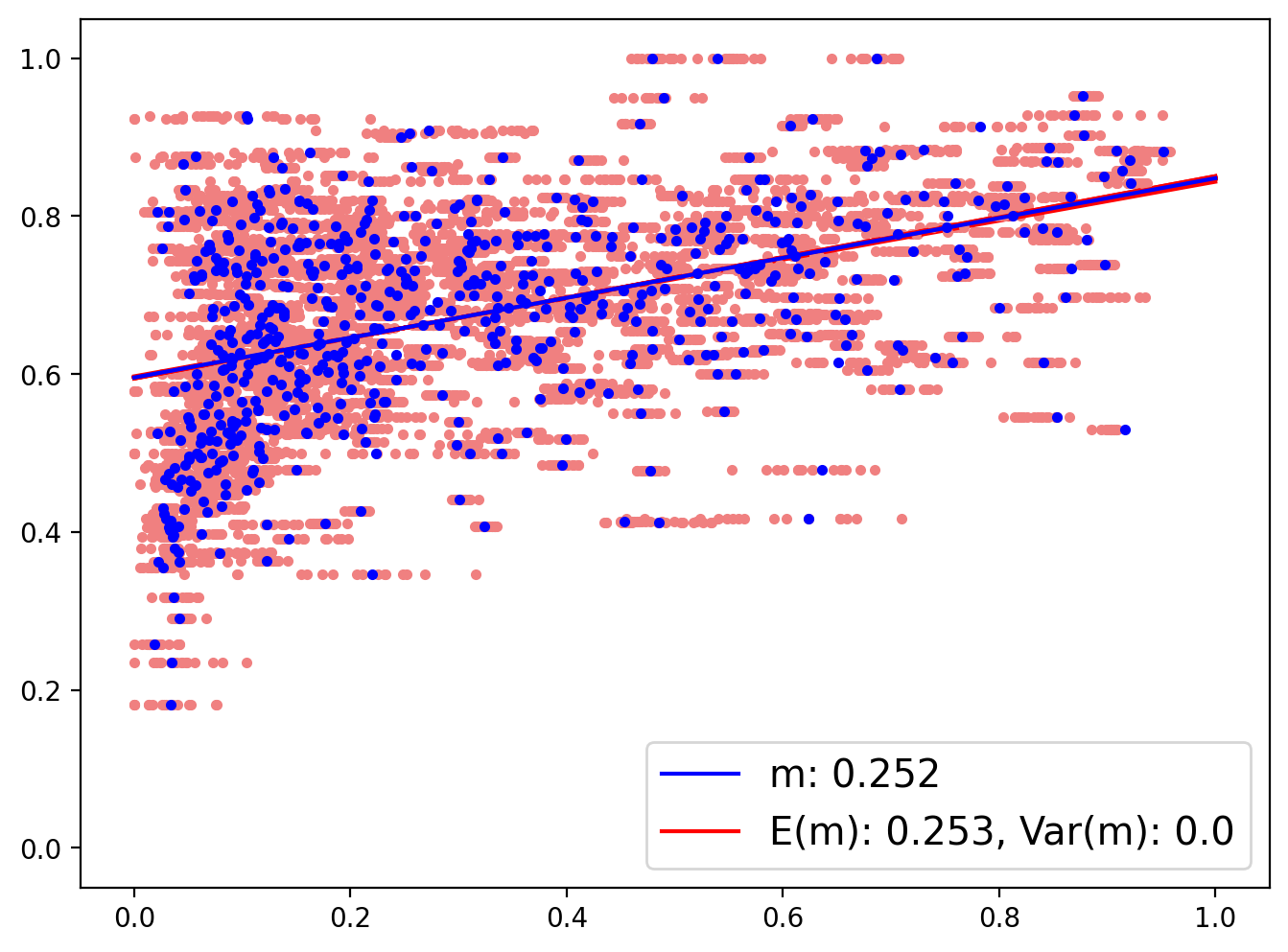}};

\node at (4,3) {\includegraphics[width=\localwidth]{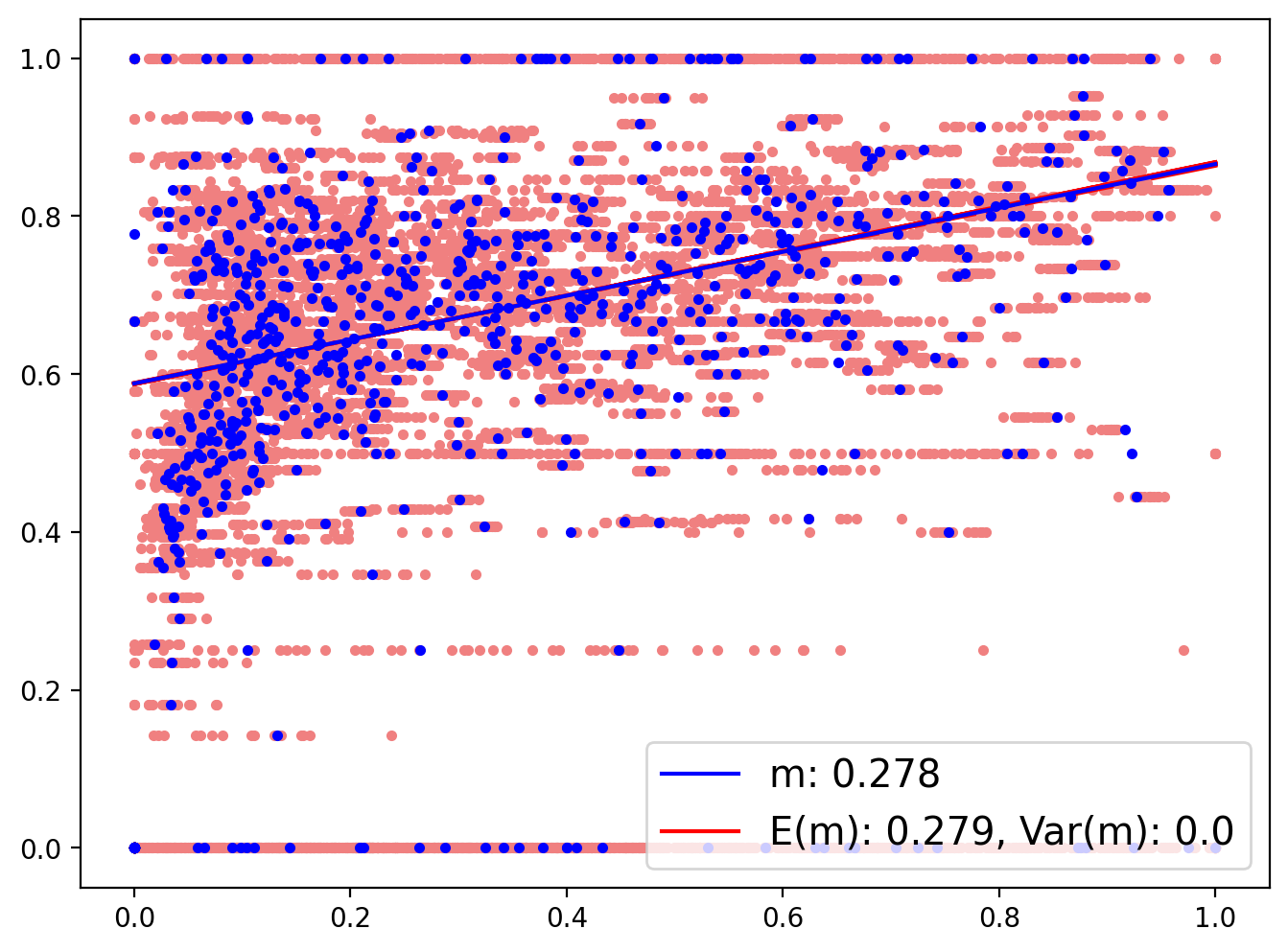}};

\node at (-4,0) {\includegraphics[width=\localwidth]{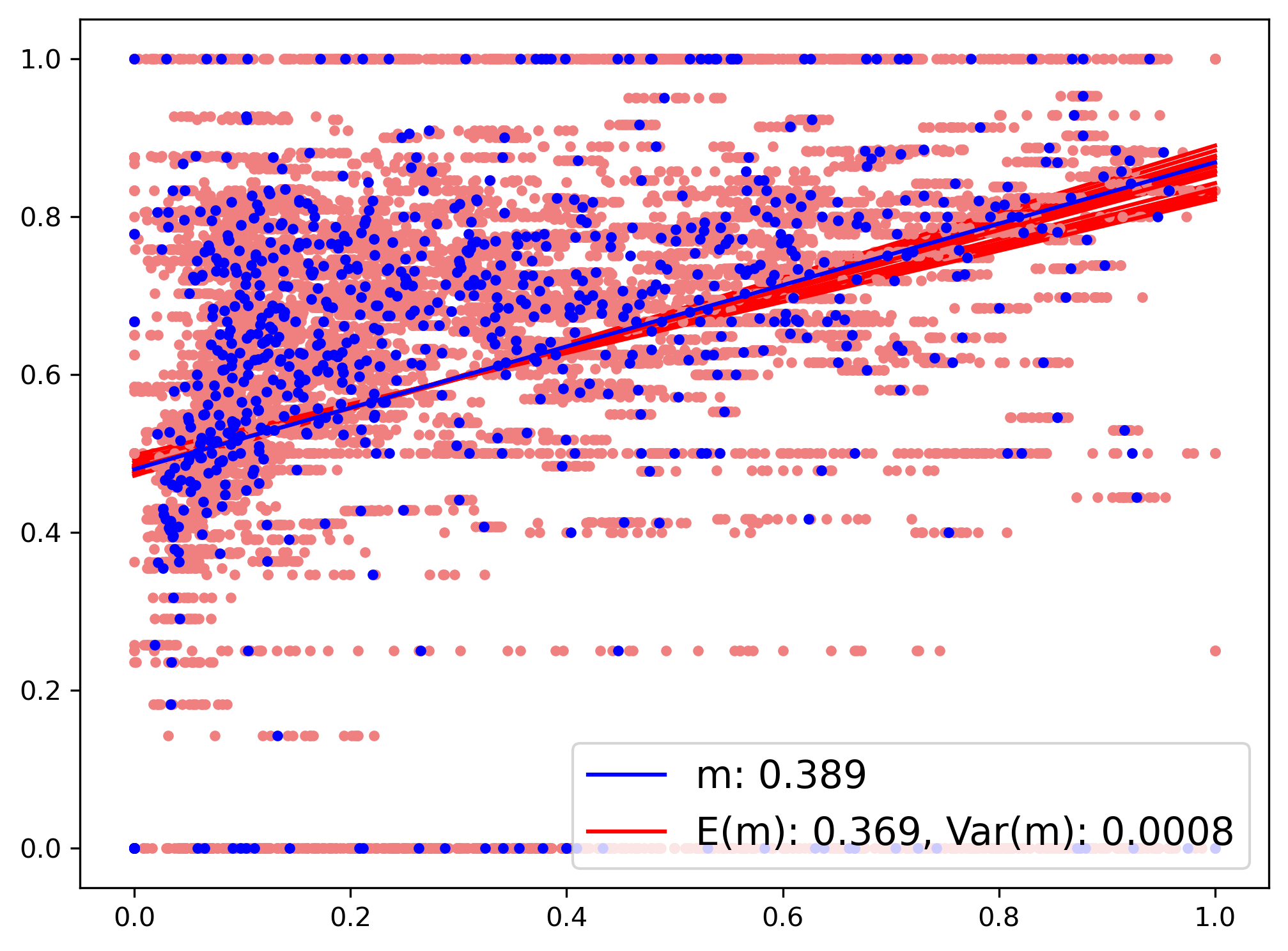}};

\node at (0,0) {\includegraphics[width=\localwidth]{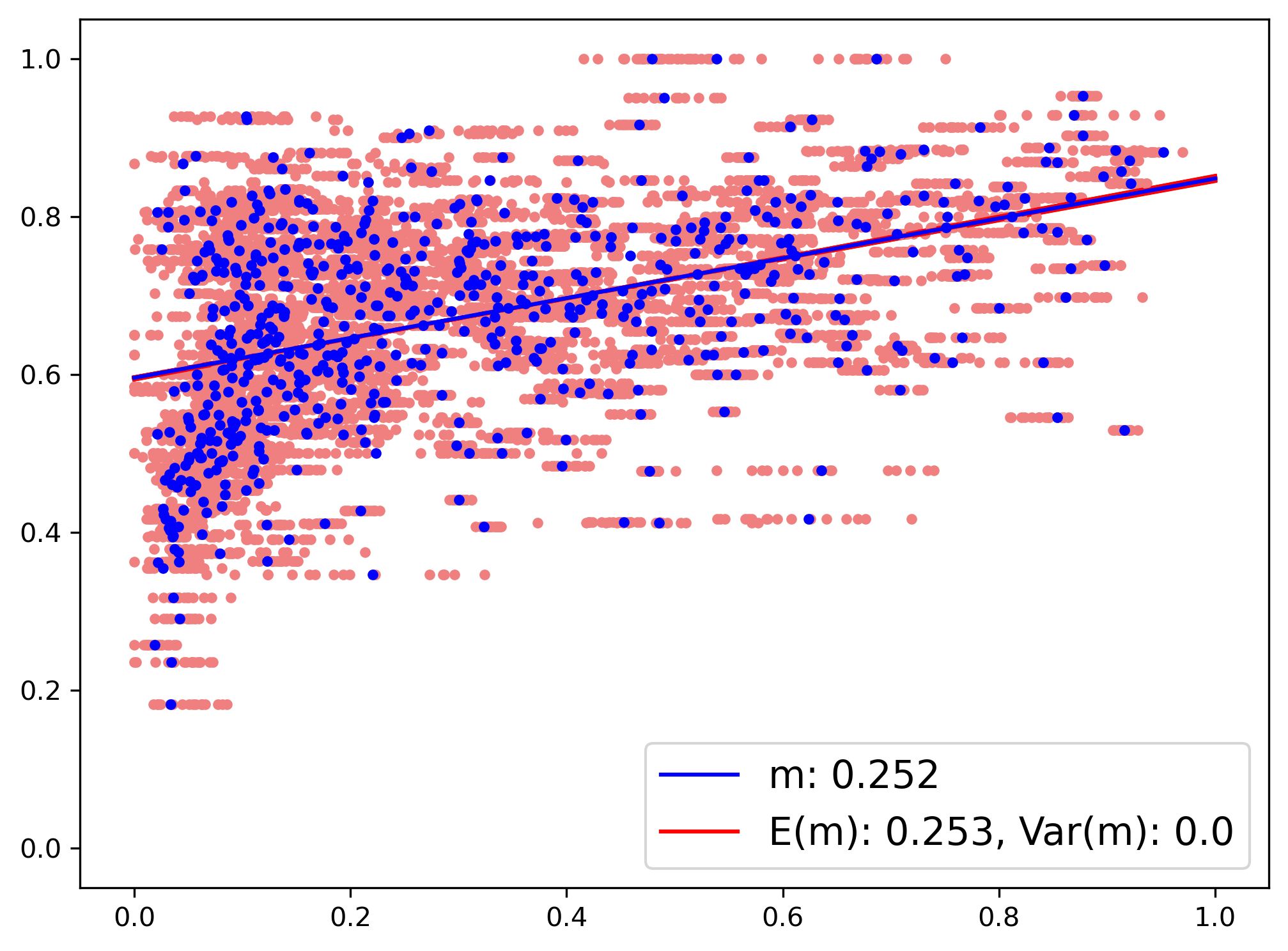}};

\node at (4,0) {\includegraphics[width=\localwidth]{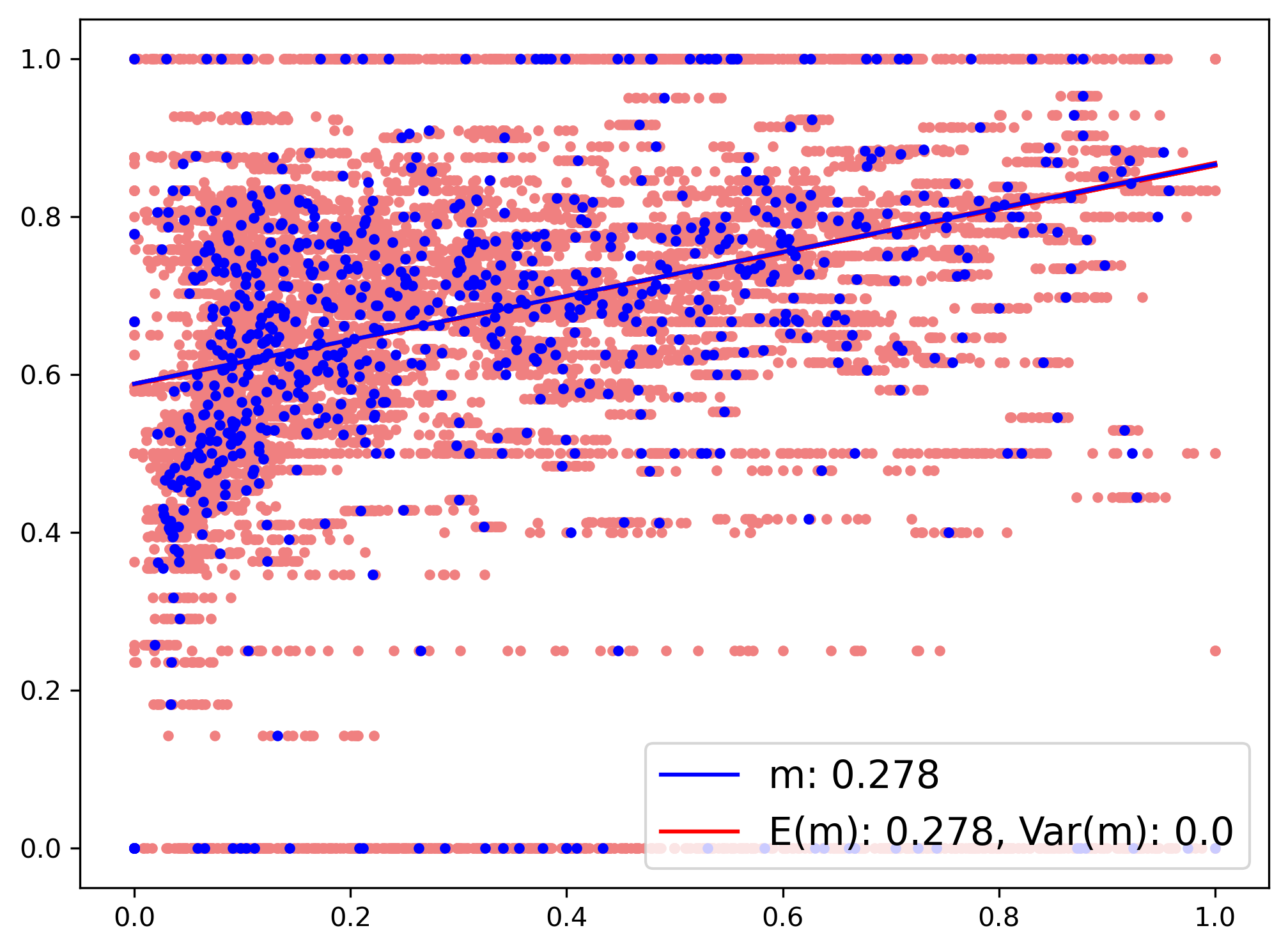}};

\node at (-4,-3) {\includegraphics[width=\localwidth]{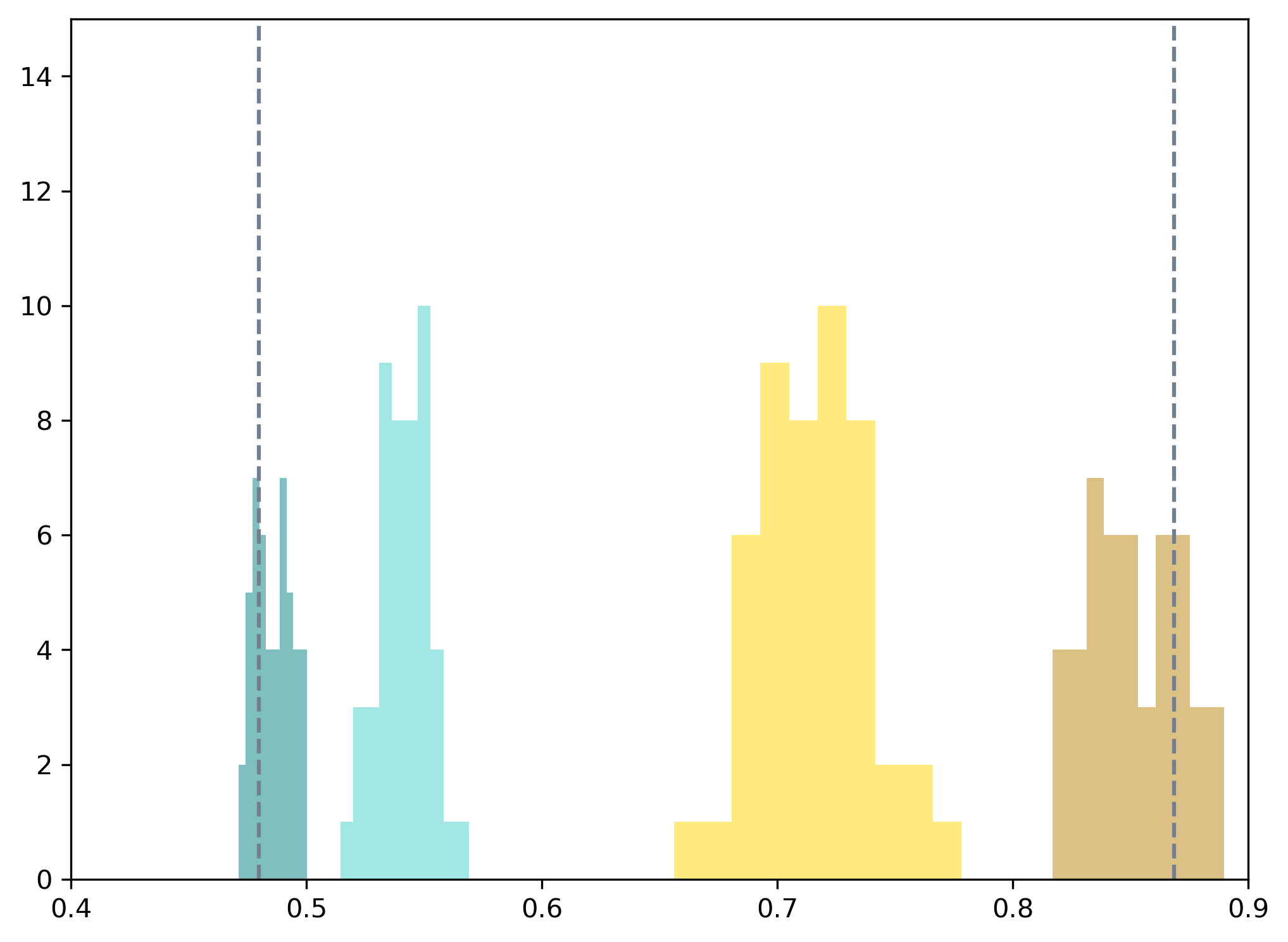}};

\node at (0,-3) {\includegraphics[width=\localwidth]{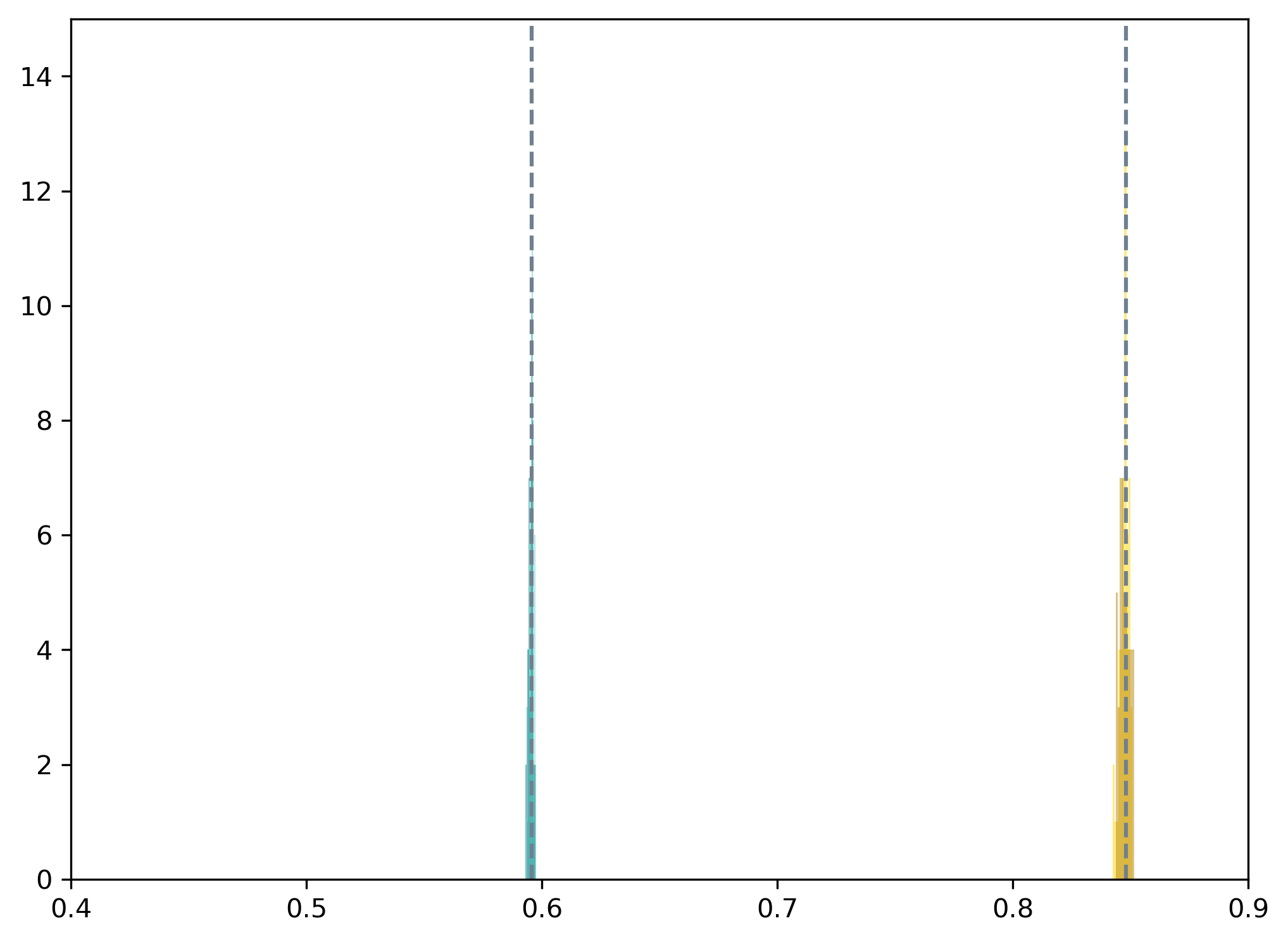}};

\node at (4,-3) {\includegraphics[width=\localwidth]{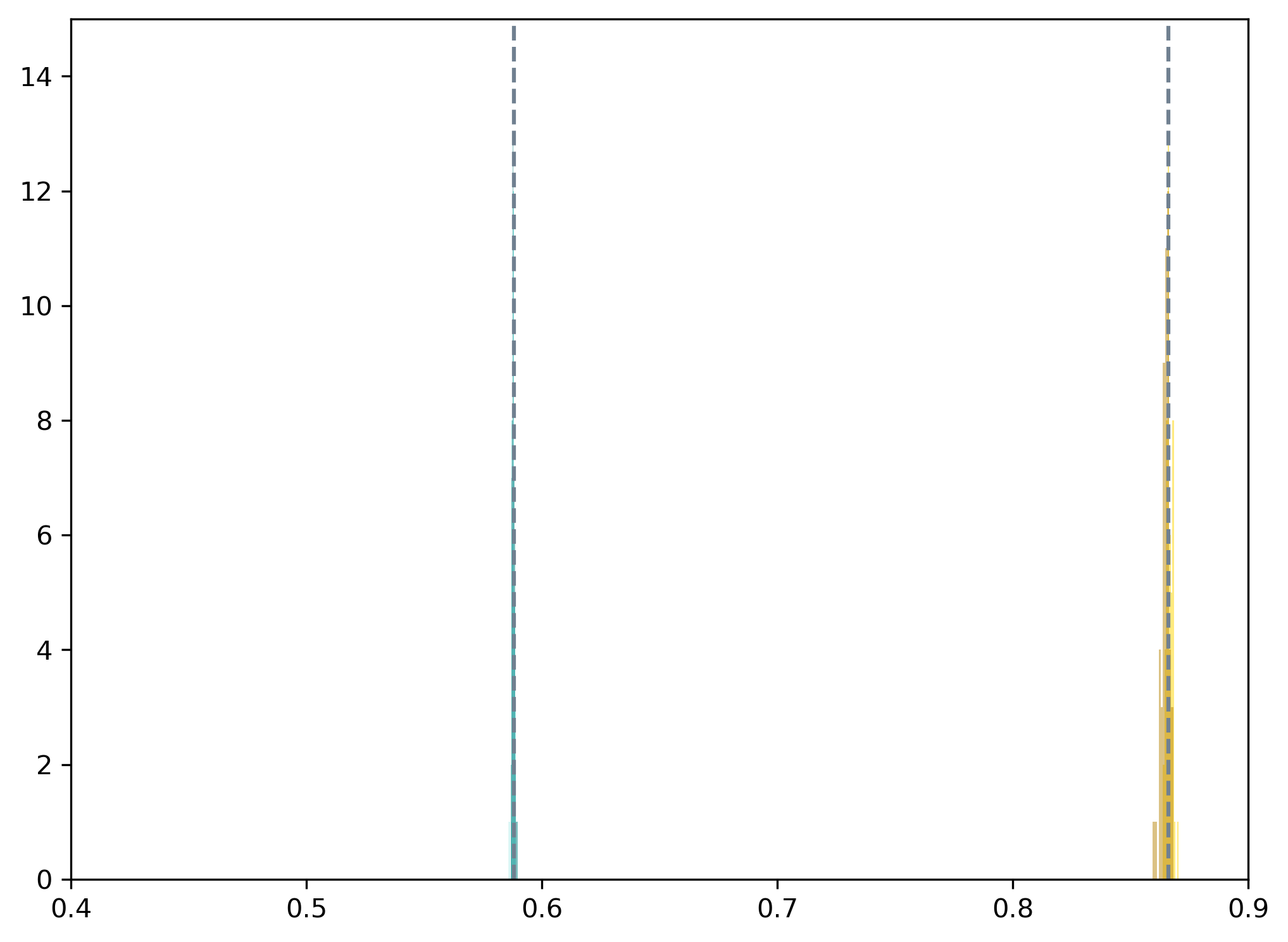}};

\draw [ultra thick] (-6,-4.75) rectangle (6,4.75);
\node at (-6.5,0) [rotate=90] {Differential privacy};

\node at (-4,-7) {\includegraphics[width=\localwidth]{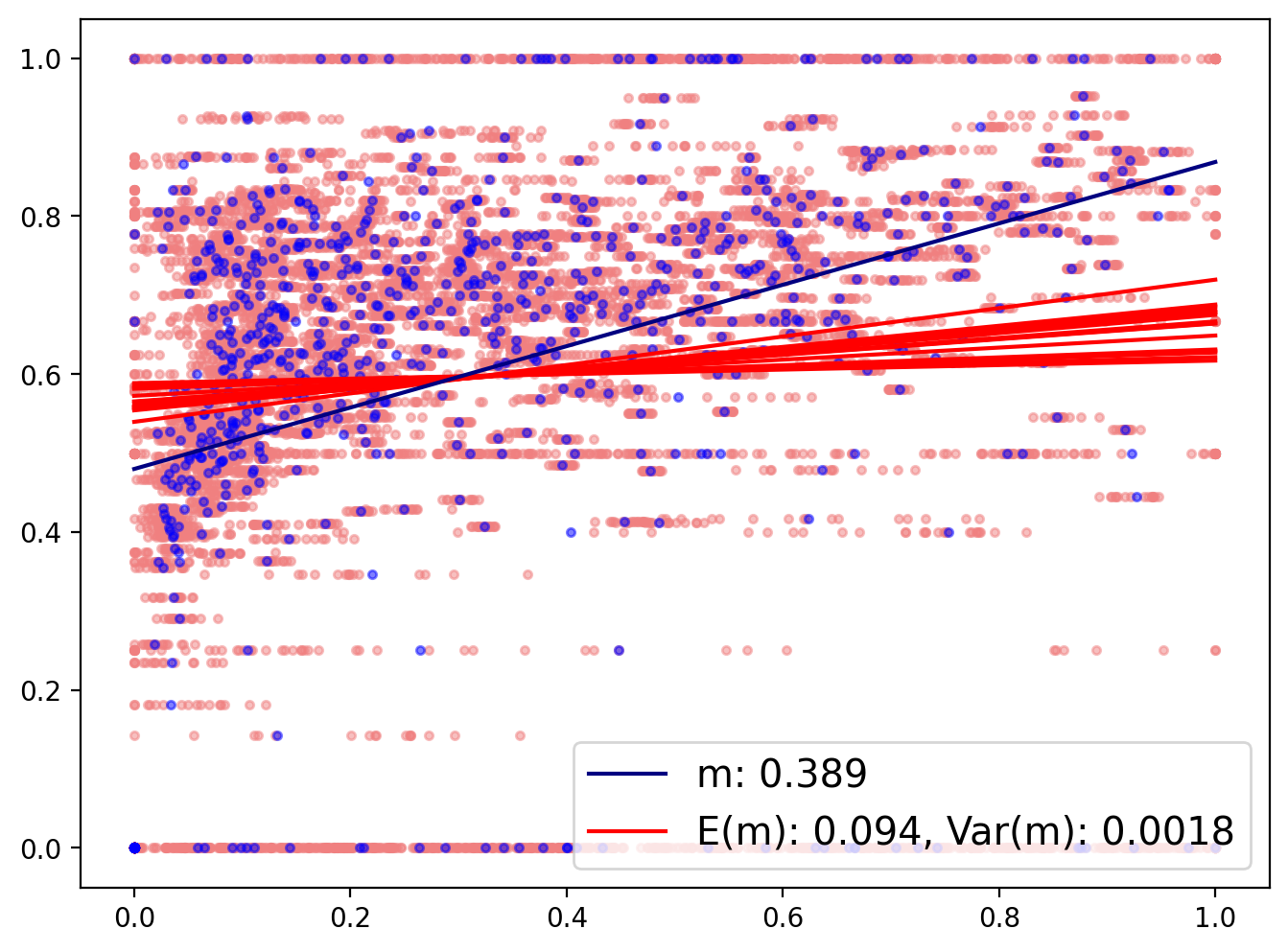}};

\node at (0,-7) {\includegraphics[width=\localwidth]{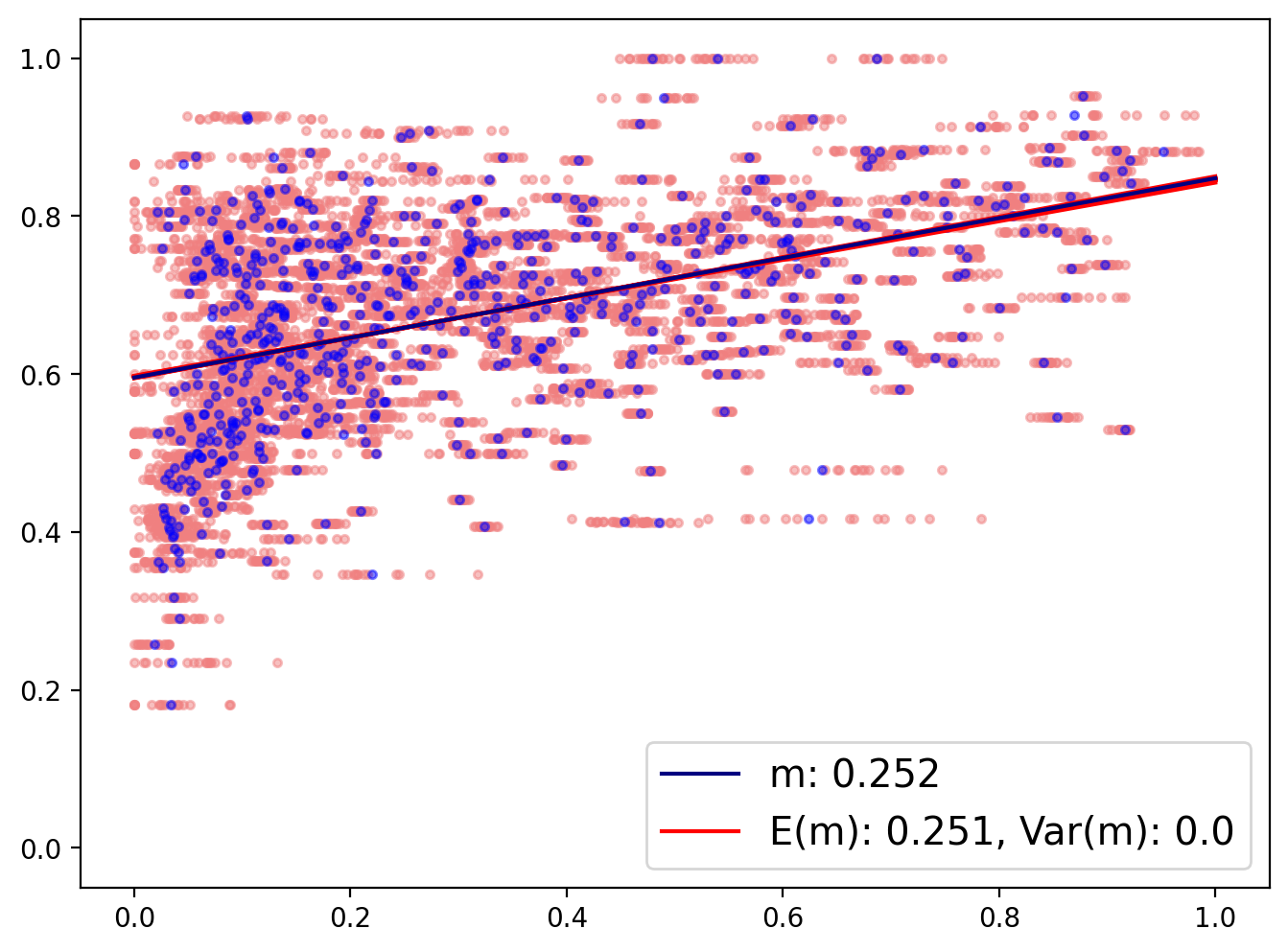}};

\node at (4,-7) {\includegraphics[width=\localwidth]{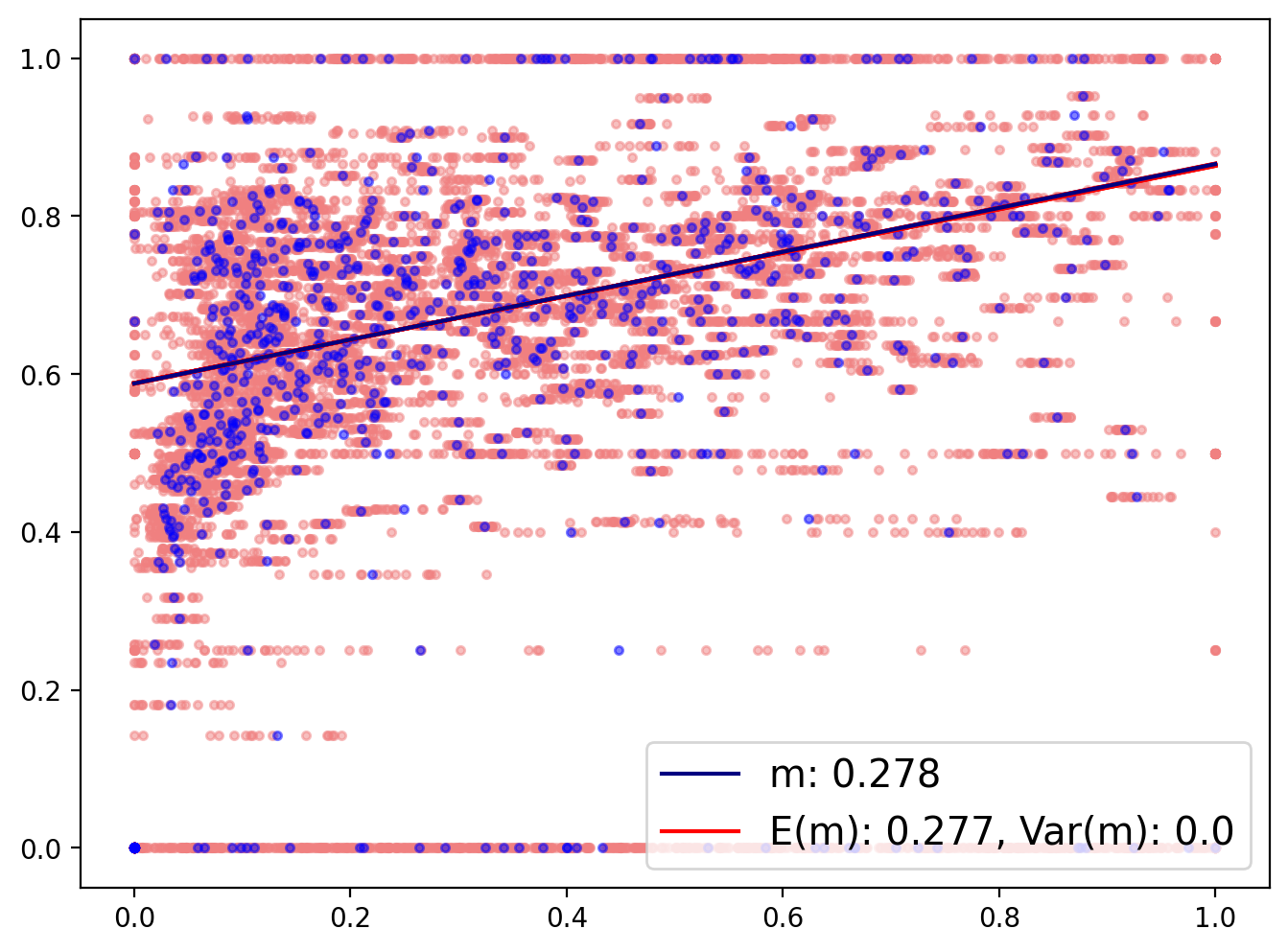}};

\draw [ultra thick] (-6,-8.75) rectangle (6,-5.25);
\node at (-6.5,-7) [rotate=90] {Gaussian};

    \end{tikzpicture}
\caption{Comparing ecological regression on un-noised data (top row) with various styles of noising. ER is re-run on data noised by differentially private \Toy (second row), and data noised by \TopDown (third row), both with $\eps=1$, equal split. The blue dots repeat the un-noised data, the pink dots show 16 runs of noised data with pink fit lines re-computed each time. Below that, the histograms show the point estimates of Latino (gold) and non-Latino (teal) support for Valdez estimated from ER on data noised by \Toy (lighter) and \TopDown (darker).
The last row contrasts the differentially private algorithms with a naive variant that adds noise  to each precinct from a mean-zero Gaussian distribution, set to match the average precinct level $L^1$ error observed in the \Toy runs (in this case, this is $\sigma = 20$). Across all of these experiments, the conclusion is striking: \TopDown performs better than \Toy and far better than a naive Gaussian variant, even without filtering precincts; if precincts are filtered or weighted, none of the noising alternatives threatens the ability to detect racially polarized voting.}
    \label{fig:er}
\end{figure}

The top row of Figure~\ref{fig:er} shows standard ER run on the precincts of Dallas County, with each precinct plotted according to its percentage of Hispanic voting age population or HVAP ($x$-axis) and the share of cast votes that went to Lupe Valdez ($y$-axis).  Strong racial polarization would show up as a fit line of high slope.
This process produces a point estimate of Hispanic support for Valdez, found by intersecting the fit line with the $x=1$ line, which represents the scenario of 100\% Hispanic population.  The point estimate of non-Hispanic support for Valdez is at the intersection of the fit line with
$x=0$.

\begin{table}[ht]
\definecolor{hisppeople}{rgb}{0.87, 0.36, 0.51}
\definecolor{blackpeople}{rgb}{1.0, 0.75, 0.0}
\definecolor{whitepeople}{rgb}{0.61, 0.87, 1.0}
    \centering
\small{
\begin{tabular}{r||c|c||c|c||c|c}
        \multicolumn{1}{c||}{} & \multicolumn{2}{c||}{All precincts (827) } & \multicolumn{2}{c||}{Filtered precincts (626)}
        & \multicolumn{2}{c}{Weighted precincts (827)}\\
        Race & this group &  complement & this group & complement & this group & complement\\
        \hline
\rowcolor{hisppeople}      Hispanic   & 0.869 & 0.480  & 0.848 & 0.596 & 0.866 & 0.588 \\
\rowcolor{blackpeople}        Black   & 0.917 & 0.518  & 0.851 & 0.620 & 0.835 & 0.595 \\
\rowcolor{whitepeople}        White   & 0.555 & 0.623  & 0.474 & 0.811 & 0.478 & 0.805
\end{tabular}

\medskip
{\small
\begin{tabular}{r|c|c||c|c||c|c||c|c}
        \multicolumn{3}{c||}{} & \multicolumn{2}{c||}{All  (827) } & \multicolumn{2}{c||}{Filtered  (626)}
        & \multicolumn{2}{c}{Weighted  (827)}\\
        Race & Algorithm & statistic &  group &  compl. &  group & compl. &  group & compl.\\
        \hline
  \rowcolor{hisppeople}        Hispanic & \Toy & mean & 0.715  & 0.541  & 0.848  & 0.595 & 0.867 & 0.588 \\
        Hispanic & \Toy  & variance & 36000 & 7000 & 250 & 43  & 160 & 19 \\
  \rowcolor{blackpeople}         Black & \Toy &  mean & 0.798 & 0.543 & 0.851 & 0.62 & 0.835 & 0.595 \\
        Black & \Toy  & variance & 39000 & 2100 & 89 & 5.9 & 25 & 2.1 \\
\rowcolor{whitepeople}         White & \Toy  & mean & 0.476 & 0.674 & 0.473 & 0.811 &  0.478 & 0.805 \\
        White & \Toy & variance & 17000
 & 8000 & 64 & 36 & 33 & 17 \\
 \hline
 \hline
  \rowcolor{hisppeople}        Hispanic & \TopDown  & mean & 0.853 & 0.485 & 0.848 & 0.595 & 0.865 & 0.587\\
        Hispanic & \TopDown  & variance  & 45000 & 6700 & 480 & 100 & 120 & 16\\
\rowcolor{blackpeople}        Black & \TopDown  & mean & 0.91 & 0.52 & 0.85 & 0.62 & 0.835 & 0.595\\
        Black & \TopDown  & variance  & 30000 & 1200 & 250 & 23 & 45 & 2.4\\
\rowcolor{whitepeople}         White & \TopDown  & mean & 0.582 & 0.607 & 0.472 & 0.81 & 0.47 & 0.804\\
        White & \TopDown  & variance  & 10000 & 3400 & 92 & 37 & 92 & 10
    \end{tabular}}}
\caption{Point estimates from ER for Dallas County in the Valdez/White primary runoff in 2018.
In the first table, estimates are made with (un-noised) VAP data from the 2010 Census.  In the {\em filtered precincts} case, precincts with fewer than 10 cast votes are excluded from the initial set of 827 precincts.  In the {\em weighted precincts} case, precincts are weighted by the number of cast votes.
The \Toy and \TopDown estimates are made from VAP data from 16 runs with $\epsilon=1$ and an $\epsilon$-budget with all levels given equal weighting.  Variance is the empirical variance over the repeated runs of the noising algorithm and is in units of $10^{-8}$, shown to two significant digits.}
\label{tab:er}
\end{table}

\subsection{Summary of Experiments}

\Toy and \TopDown were both run on the full Texas reconstruction from 2010. We plotted Dallas County votes from three contests: votes for Obama for president in 2012 general election, votes for Valdez for governor in the 2018 Democratic Party primary runoff, and votes for Chevalier for comptroller in the 2018 general election.  We chose these contests because in each, ER finds evidence of strong racially polarized voting when using published 2010 census data. All three contests gave similar findings; we'll choose the Valdez runoff contest as our focus here.

For both \Toy and \TopDown, we vary how we handle the inclusion of small precincts in the ecological regression. The options are All (every precinct is a data point in the scatterplot, all weighted equally); Filtered (only including precincts with at least 10 votes cast in that election); or Weighted (weighting the terms in the objective function in least-squares fit by number of votes cast). Filtering and weighting are done using the exact number of cast votes, not the differentially private precinct population totals, which is realistic to the use case.

For each noising run we have a block- or precinct-level matrix, $\hat{M}$ of noised counts, with height $b$, the number of geographic units (blocks or precincts), and width $c$, the number of attributes for which there are counts recorded. We also have a corresponding matrix $M$ of un-noised counts.  We can compute the $L_1$ error by summing over the absolute value of every entry in $M - \hat{M}$.
\Toy and \TopDown were run 16 times for each configuration.
Let $E_{avg}$ be the average $L_1$ error across noising runs.

If we add {\em Gaussian} noise to each count instead, the expected $L_1$ error is $\sum_{i,j} E[|X_{i,j}|]$, where $X_{i,j} \sim \mathcal{N}(0,\,\sigma^{2})$.  This is the half-normal distribution, so $E[|X_{i,j}|] = \frac{\sigma \sqrt{2}}{\sqrt{\pi}}$. We rearrange to find the standard deviation $\sigma = \frac{E_{avg} \sqrt{\pi}}{bc\sqrt{2}}$ that defines the Gaussian distribution (with $\mu = 0$), so that adding a random variable drawn from it to each unit count will produce an expected $L^1$ error matching the average $E_{avg}$ observed across the runs.

\subsection{The role of small precincts}

Practitioners who use ER have raised two questions regarding the effect of differential privacy: (1) How robust will the estimate be after the noising? (2) Will noising diminish the estimate of candidate support from a minority population?
We analyzed the effects of \TopDown and \Toy on the 2018 Texas  Democratic primary runoff election, where Lupe Valdez was a clear minority candidate of choice in Dallas county.\footnote{We also examined the general elections for President in 2012 and Comptroller in 2018, with similar findings.}

We begin by observing that of the 827 precincts in Dallas County, 201 have fewer than 10 cast votes from that election day---in fact, 99 precincts recorded zero cast votes.  These precincts are a big driver of instability under DP.  This is not surprising; percentage swings are much higher in small numbers even if the noise injected might be low. However, down-weighting these small precincts makes the estimate almost always agree with the un-noised estimate. Specifically, we assign weights to the precincts equivalent to the number of total votes in the precinct. Figure \ref{fig:er} shows how the estimates vary by run type and data treatment.

\section{Conclusion}

The central goal of this study has been to take the concerns of redistricting practitioners seriously and to investigate potential destabilizing effects of \TopDown on the status quo.  A second major goal is to make recommendations, both to the Disclosure Avoidance team at the Census Bureau and to the same practitioners---the attorneys, experts, and redistricting line-drawers in the field.
Texas generally, and Dallas County in particular, was selected because it has been the site of several interesting Voting Rights Act cases in the last 20 years.\footnote{This is a large county with considerable racial and ethnic diversity.  Follow-up work will consider smaller and more racially homogeneous localities.}

Our top-line conclusion is that, at least for the Texas localities and election data we examined, \TopDown performs far better than more naive noising in terms of preserving accuracy and signal detection for election administration and voting rights law.
Perhaps more importantly, we have created an experimental apparatus to help other groups conduct independent analyses.

This work has led us to isolate several elements of common redistricting practice that lead to higher-variance outputs and more error under \TopDown.  The first example is the common use of a full precinct dataset, with no population weighting, in running racial polarization inference techniques.  The second major example is the use of the smallest available units, census blocks, for building districts of all sizes, with no particular priority on intactness for larger units of Census geography.  In both cases, we find that these were already likely sources of silent error. Filtering small precincts (or, better, weighting by population) and building districts that prioritize preserving whole the largest units that are suited to their scale are two examples of simple updates to redistricting practice. Besides being sound on first principles, these adjustments can insulate data users from DP-related distortions and help safeguard the important work of fair redistricting.

\newpage
\bibliography{bibliography}

\appendix

\section{\Toy and \TopDown}
\label{app:ToyTop-description}

\Toy is described in Algorithm~\ref{alg:Toy}. It uses the {\em Laplace distribution}  $\Lap(b)$ with scale parameter $b$, i.e., the  probability distribution over $\R$ with mean zero and probability density function $\Pr[L] = \frac{1}{2b}e^{-|L|/b}$. It has variance $2b^2$.
\TopDown uses the {\em geometric} distribution, a discretized version of the Laplace distribution with integer support.

The inputs to \TopDown are as follows. $A_{H,T} = \{a_{h,t}\}_{h\in H, t \in T}$, where $a_{h,t}$ is the number of people in $h$ of type $t$; $W = (Q_1,\dots, Q_{|W|})$ is a \emph{workload} consisting of a collection of histograms $Q$; $\eps=(\eps_1,\dots,\eps_d)$ is a hierarchical allocation of the privacy budget, with $\eps_\ell>0$ at each level; $B:W \to [0,1]$ with $\sum_{Q\in W} B(Q)=1$ is a probability vector describing the relative privacy budget on each histogram in the workload; \emph{invariants} $V$; and \emph{structural inequalities} $S$.
We write $\bm{a}_h = \{a_{h,t}\}_{t\in T}$ (and $\bm{\alpha}_h$ analogously). For a query $q$, we write $q(\bm{a}_h) = \sum_{t \in q} a_{h,t}$ (and $q(\bm{\alpha}_h)$ analogously).

In the first stage (lines 2-5), a geometric random variable is added to the raw counts $a$ to produce noised counts $\hat a$.  In the second stage (lines 6-8), the noised counts are adapted to the nearest integer values that meet a collection of equality and inequality conditions.  These equalities and inequalities, over the real numbers, describe a convex polytope; therefore the post-processing can be thought of geometrically as a closest-point projection to the integer points in the convex body under $L^2$ distance.

The noising stages of both \Toy and \TopDown are $\eps$-differentially private for $\eps = \sum_{\ell =1}^d \eps_\ell$.
In \Toy, this stage can be viewed as generating a single histogram at each level $\ell$ using budget $\eps_\ell$. Following the Census Bureau, we use bounded differential privacy, wherein the global sensitivity of histogram queries is 2.
In \TopDown, the budget at level $\ell$ is further divided among the $|W|$ histograms $Q$ in the workload, each receiving $B(Q)\eps_\ell$ of the budget.
Because \Toy's post-processing is data independent, \Toy is $\eps$-DP. \TopDown's post-processing is not data independent: the invariants and structural inequalities may depend on the original data.

\begin{algorithm}[H]
    \caption{$\TopDown$, based on \cite{TopDown}}
    \label{alg:TopDown}
    \begin{algorithmic}[1] %
        \Procedure{$\TopDown$}{$A_{H,T}$, $\eps_1,\eps_2,\dots,\eps_d$, $W$, $B$, $V$, $S$}
            \For{$h \in H$, $Q\in W$, $q\in Q$}
                \State $\beta \gets \exp(-B(Q)\cdot \eps_{\ell(h)}/2)$
                \State $G_{h,q} \gets \mathrm{Geom}(\beta)$ \Comment{See \cite{topdown_geometric}}
                \State $\ahat_{h,q}\gets q(\bm{a}_h) + G_{h,q}$ \Comment{Geometric mechanism with }
                \Statex
                \continueComment{sensitivity 2, budget $B(Q)\cdot\eps_{\ell(h)}$}
            \EndFor
            \For{$\ell= 1,\dots,d$}
                \State Compute hierarchically-consistent  \label{alg:line:hard:first} \continueComment{\(\triangleright\) A sophisticated heuristic algorithm}
                \Statex\hspace{\algorithmicindent}\hspace{\algorithmicindent}non-negative integers $\{\alpha_{h,t}\}_{h \in H_\ell,t \in T}$ \continueComment{out of scope for this work}
                \Statex\hspace{\algorithmicindent}\hspace{\algorithmicindent}minimizing $\sum_{h \in H_\ell} \sum_{q\in W_\ell} \left(q(\bm{\alpha}_{h}) - \ahat_{h,q} \right)^2$,
                \Statex\hspace{\algorithmicindent}\hspace{\algorithmicindent}subject to the invariants: $v^*(\bm{\alpha}_h) = v^*(\bm{a}_h)$ for all $h\in H_\ell$, $v \in V$
                \Statex\hspace{\algorithmicindent}\hspace{\algorithmicindent}and structural inequalities: $s(\bm{\alpha}_h, \bm{a}_h) \le 0$ for all $h\in H_\ell$, $s \in S$
            \EndFor
            \State\Return $\{\alpha_{h,t}\}_{h\in H,t\in T}$
        \EndProcedure
    \end{algorithmic}
\end{algorithm}

\begin{algorithm}
    \caption{$\Toy$}
    \label{alg:Toy}
    \begin{algorithmic}[1] %
        \Procedure{$\Toy$}{$A_H = \{a_h\}_{h\in H},  \eps_1,\eps_2,\dots,\eps_d$} \Comment{(Single attribute)} \label{line:Toy:single}
            \For{$h \in H$}
                \State $L_h \sim \Lap(2/\eps_{\ell(h)})$
                \State $\ahat_h \gets a_h + L_h$   \Comment{Laplace mechanism with sensitivity 2, budget $\eps_{\ell(h)}$} \label{alg:toy-single:line:laplace}
            \EndFor
            \For{$\ell= 1,\dots,d$}
                \State Compute hierarchically consistent $\{\alpha_h\}_{h \in H_\ell}$
                \Statex\hspace{\algorithmicindent}\hspace{\algorithmicindent}minimizing $\sum_{h \in H_\ell} (\alpha_h - \ahat_h)^2$
            \EndFor
            \State\Return $\{\alpha_h\}_{h\in H}$
        \EndProcedure
        \Statex %

        \Procedure{$\mathsf{MultiAttr}\Toy$}{$A_{H,T} = \{a_{h,t}\}_{h\in H, t \in T}, \eps_1,\eps_2,\dots,\eps_d$}  \label{line:Toy:multi}
            \For{$h \in H$, $t \in T$}
                \State $L_{h,t} \sim \Lap(2/\eps_{\ell(h)})$
                \State $\ahat_{h,t} \gets a_{h,t} + L_{h,t}$ \Comment{Laplace mechanism with sensitivity 2, budget $\eps_{\ell(h)}$} \label{alg:toy-multi:line:laplace}
            \EndFor
            \For{$\ell= 1,\dots,d$}
                \State Compute hierarchically consistent
                \Statex \hspace{\algorithmicindent}\hspace{\algorithmicindent}(optionally, non-negative) $\{\alpha_{h,t}\}_{h \in H_\ell,t \in T}$
                \Statex\hspace{\algorithmicindent}\hspace{\algorithmicindent}minimizing
                $\sum_{h\in H_\ell, t \in T} \left(\alpha_{h,t} - \ahat_{h,t} \right)^2$
            \EndFor
            \State\Return $\{\alpha_{h,t}\}_{h\in H,t\in T}$
        \EndProcedure
    \end{algorithmic}
\end{algorithm}

\section{Detailed materials and methods}
\label{app:methods}

\subsection{Primary data sources}

2010 US Census demographic data was downloaded using the Census API, and the 2010 census block, block group, and tract shapefile for Dallas County were downloaded from the US Census Bureau’s TIGER/Line Shapefiles.
For our VRA analysis, we obtained both statewide election results and a statewide precinct shapefile from the Texas Capitol Data Portal, which we then trimmed to the precincts within Dallas County.\footnote{Data comes from \href{https://data.capitol.texas.gov/topic/elections}{data.capitol.texas.gov/topic/elections} and \href{https://data.capitol.texas.gov/topic/geography}{data.capitol.texas.gov/topic/geography}.}%

We use a person-level dataset obtained by applying a reconstruction technique to the block-level data from Texas from the 2010 Census.\footnote{A team led by data scientist and journalist Mark Hansen at Columbia, including Denis Kazakov, Timothy Donald Jones, and William Reed Palmer, designed an algorithm to solve for the detailed data, which we describe in this section. Code is available upon request \cite{Kazakov}.}
The reconstructed microdata records contain block-level sex, age, ethnicity, and race information consistent with a collection of tables from 2010 Census Summary File 1.
We note that this reconstruction follows the same strategy used by the Census Bureau itself as the first step of its reidentification experiment \cite{Hawes}, based on \cite{DinurNissim}.

The reconstructed data is far from perfect.
Unlike the Bureau, we do not have access to the ground truth data needed to quantify the errors.
The Bureau's own reconstruction experiment  reconstructed  $46\%$ of entries exactly, plus an additional 25\% within $\pm1$ year error in age \cite{Hawes}.
We note that our reconstructed data contains no household information, because this was not present in the tables used in the constraint system. This is significant because the \TopDown configurations for the US Census Bureau's 2010 Demonstration Data Products~\cite{2010-ddp} include household-based workload queries and invariants.

\subsection{\TopDown configuration}
\label{sec:topdown-config}

The exact configuration files and code for all the runs are available in this paper's accompanying repository~\cite{our-github}.
The \TopDown code used for this paper was modified from the publicly available demonstration release of the US Census Bureau's Disclosure Avoidance System 2018 End-to-End test release \cite{End-to-end}. The input data fed to the algorithm was obtained by restructuring the reconstructed 2010 block-level Texas microdata into the 1940s IPUMs data format. Most importantly, the reconstructions allowed for 63 distinct combination of races whereas the End-to-End release only allows for 6 races, so all multi-racial entries were re-categorized as Other in our \TopDown runs.

Because \TopDown's post-processing is done level by level, the noisy counts in Dallas County do not depend on the noisy counts at the tract-level or below in counties other than Dallas. We modified the census reconstructed data to focus on Dallas county and minimize the computation time spent processing the other 253 counties in Texas. Specifically, for every non-Dallas county, we placed all of the population into a single block.

We do not enforce certain household invariants that the Census Bureau is planning to enforce, and our workload omits household queries that are used in Census's demonstration data products.
Our choice to omit household queries and invariants is result of our use of reconstructed 2010 census microdata which does not include household information.
We did perform additional runs with household invariants and queries using crude synthetic household data, the results of which are available in the data repository accompanying this paper \cite{our-github}. In those runs, the population in each block was grouped into households of size 5 with at most one group smaller than 5.
Ultimately, we focused on the experiments that did not require synthetic household data.

The \TopDown runs without the household workload or invariants use a workload consisting of two histograms: $Q_{detailed}$ and $Q_{va,eth,race}$ with 10\% and 90\% of the budget respectively. (The additional runs with households includes an additional households and group quarters histogram in the workload assigned 22.5\% of the budget, leaving 10\% and 67.5\% for $Q_{detailed}$ and $Q_{va,eth,race}$ respectively.)
The End-to-End \TopDown code reports a differentially private estimate of the $L^1$ error with $\eps = 0.0001$ not included in privacy budget specified elsewhere in the configuration file and discussed elsewhere in this paper.

\section{District fragmentation}
\label{app:frag-empirical}

\begin{algorithm}
    \caption{$\Greedy$}
    \label{alg:Greedy}
    \begin{algorithmic}[1] %
        \Procedure{$\Greedy$}{$H,k$}
            \If{$k=1$}
                \State Return $H$
            \EndIf
            \State $N \gets \floor{|H_d|/k}$, $D \gets \emptyset$, $h^* \gets h_1$
            \While{$N >0$}
                \State For $h^*$ and $D$, let $S(h^*, D)$ be the set of
                \Statex\hspace{\algorithmicindent}\hspace{\algorithmicindent}children $h$ of $h^*$ that are disjoint from $D$.
                \While{$\exists h \in S(h^*,D) : |h| \le N$} \label{alg:greedy:1}
                    \State $D \gets D\cup h$  \Comment{Associating $h$ with the blocks descendent from it}
                    \State $N \gets N - |h|$
                \EndWhile
                \State Pick $h^*\in S(h^*,D)$ \label{alg:greedy:2}
            \EndWhile
            \Return $D$
        \EndProcedure
    \end{algorithmic}
\end{algorithm}

\begin{algorithm}
    \caption{$\Square$}
    \label{alg:Square}
    \begin{algorithmic}[1] %
        \Procedure{$\Square$}{$H,k$}
            \State $s_d \gets \sqrt{|H_d|/k}$ \Comment{Side length in blocks  of the district }
            \State $S_d \gets \sqrt{n_1\cdot n_2\cdots n_{d-1}}$ \Comment{Side length in blocks of the region}
            \State Sample $i,j \in \{1,\dots,S_d-s_d+1\}$ uniformly at random
            \State \Return $D_{i,j}$, the square district with top left corner at $(i,j)$
        \EndProcedure
    \end{algorithmic}
\end{algorithm}

In Section~\ref{sec:fragmentation-definition}, we defined the fragmentation score and its relationship to error variance for \Toy, and analyzed the expected fragmentation score of districts produced by different district drawing algorithms.
Now we apply \TopDown to examine the relationship between a district's population error and geometry, as captured by the fragmentation score.

We fix the a total budget and an equal allocation across levels: $0.2=\eps_{2}=\eps_{3}=\eps_{4}=\eps_{5}=\eps_{6}$, as in Table~\ref{tab:budget-splits}.
(We do not need to noise the nation because we are focusing on Texas; we do need to noise Texas even though its total population is invariant, because its population by race is allowed to vary.)
We apply \ReCom to build districts out of  tracts, block groups, and blocks---all of which are part of the census hierarchy---and add a realistic variant that builds from whole \emph{precincts}. These are about the same size as block groups and are more commonly used in redistricting.

\begin{figure}[ht]
    \centering
    \includegraphics[width=0.9\textwidth]{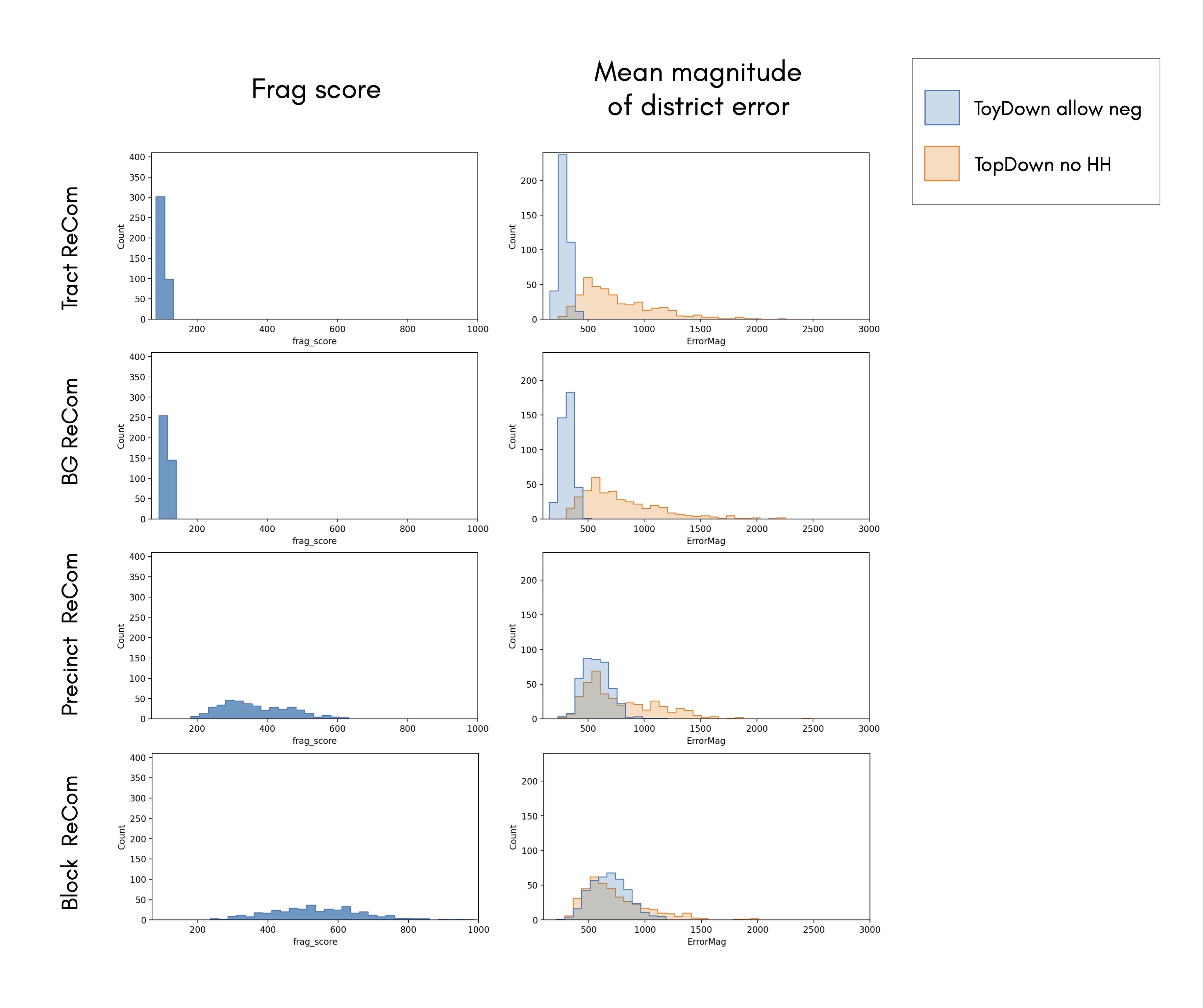}
\caption{Do the building-block units of districts matter? Histograms of fragmentation score (left column) and mean error magnitude (right column) are shown across four district-drawing algorithms that prioritize compactness.  (Dallas County, $k=4$.)  We see that using larger units leads to significantly lower fragmentation and correspondingly low district-level error in \Toy, but the advantage erodes when we pass to \TopDown.}
    \label{fig:dist-level-error-mag}
\end{figure}

Figure~\ref{fig:dist-level-error-mag} plots the data from our experiments.
Each of the 12 histograms displays 400 values, one for each district drawn by the specified district-drawing algorithm.
The histograms on the left plot the fragmentation score of each district;
the histograms on the right plot the mean observed district-level population error magnitude over 16 executions of the specified hierarchical noising algorithm.

The size of the constituent units is observed to have a controlling effect on the fragmentation score, as expected.  As we would expect,  this carries over to the simplest \Toy (allowing negativity).  (Note that since the error has zero mean, higher variance drives up the mean magnitude of error.)  But the choice of base units makes far less difference by the time we move to full \TopDown.
These observations are consistent, again, with a strong similarity across spatially nearby units.  All four kinds of \ReCom will tend to produce compact, squat districts whose units are more closely geographically proximal than would be observed with disconnected or elongated shapes.  Random noise is uncorrelated, but the post-processing effects can be highly spatially correlated because of spatial relationships in the underlying counts by race, ethnicity, and voting age.

\section{Robustness of noisy ER}
\begin{figure}[ht]
    \centering
\begin{tikzpicture}

\node at (0,5) {\footnotesize Ecological regression};
\node at (-2,4.5) {\footnotesize equal split};
\node at (.1,4.5) {\footnotesize block-heavy};
\node at (2.15,4.5) {\footnotesize tract-heavy};

\node at (-3.7,3) [rotate=90] {$\eps=0.5$};
\node at (-3.7,1) [rotate=90] {$\eps=2$};

\node at (0,2) {\includegraphics[height=1.6in]{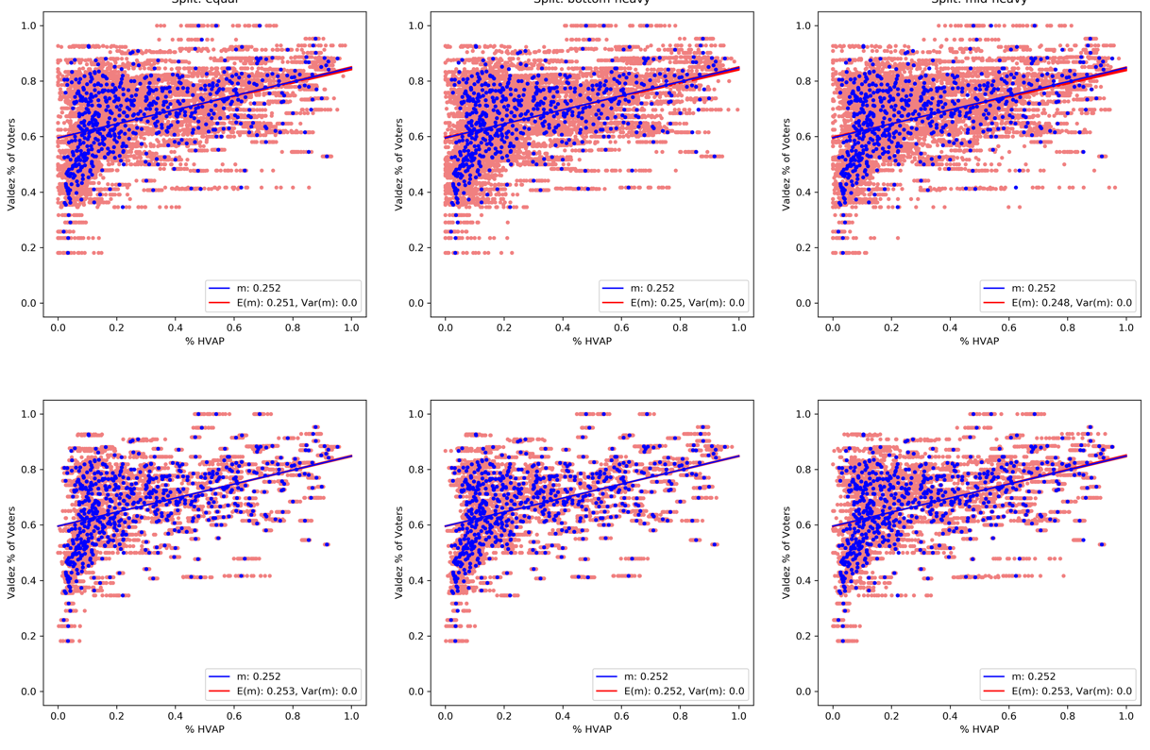}};
\node at (7,2) {\includegraphics[height=1.6in]{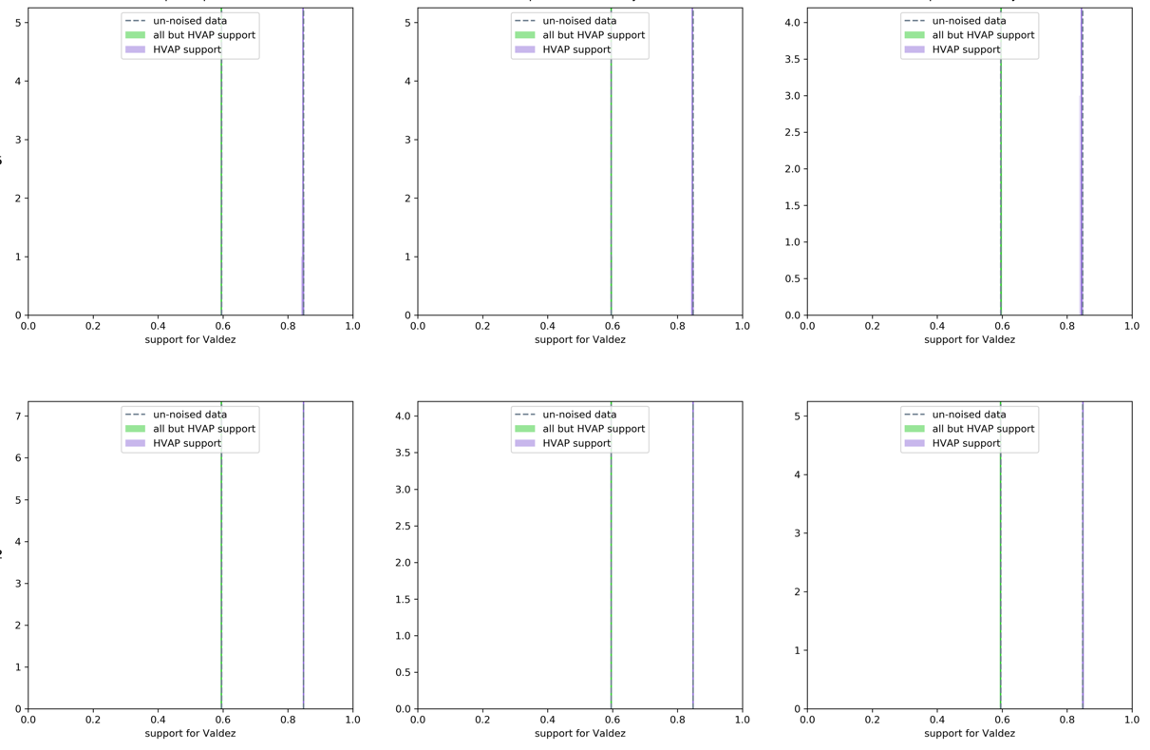}};

\node at (7,5) {\footnotesize Point estimates};
\node at (5,4.5) {\footnotesize equal split};
\node at (7,4.5) {\footnotesize block-heavy};
\node at (9.15,4.5) {\footnotesize tract-heavy};

  \end{tikzpicture}

    \caption{Ecological regression
for the Valdez-White runoff election with $\eps=.5$ and $\eps=2$ and three different budget allocations, together with corresponding point estimates for Latino and non-Latino support for Valdez, with small precincts filtered out as in Figure~\ref{fig:er}. Findings stay remarkably stable.}
    \label{fig:manysplits}
\end{figure}

Figure~\ref{fig:manysplits} extends the findings from Figure~\ref{fig:er} with more splits and allocations, showing that as long as small precincts are filtered out, ecological regression for RPV analysis in Dallas County is robust to changes in the allocation of the privacy budget across the levels of the hierarchy and the total privacy budget for \TopDown. The corresponding plots for \Toy are essentially indistinguishable. (ER with precincts weighted by population is similarly robust.)

\end{document}